\documentclass[draftcls, onecolumn, 10pt]{IEEEtran}
\usepackage{amssymb, amsmath, graphicx, paralist, subfigure}

\begin{document}
\title{Distributing the Kalman Filter for\\ Large-Scale Systems}
\author{Usman~A.~Khan,~\IEEEmembership{Student Member,~IEEE,}
Jos\'e~M.~F.~Moura,~\IEEEmembership{Fellow,~IEEE}\\
            Department of Electrical and Computer Engineering\\
            Carnegie Mellon University\\
            5000 Forbes Ave,
            Pittsburgh, PA 15213\\
            \{ukhan, moura\}@ece.cmu.edu\\
            Ph: (412)268-7103 Fax: (412)268-3890
\thanks{This work was partially supported by the DARPA DSO Advanced Computing and Mathematics Program
Integrated Sensing and Processing (ISP) Initiative under  ARO
grant \#~DAAD 19-02-1-0180, by NSF under grants \#~ECS-0225449
and~\#~CNS-0428404, and by an IBM Faculty Award.}
%\thanks{}% <-this % stops a space
%\thanks{U. Khan is with the Carnegie Mellon University.}
}

\markboth{IEEE Transactions on Signal Processing}{Khan \MakeLowercase{\textit{et al.}}: Kalman Filters with Distributed Models for Large-Scale Systems}
\maketitle

\begin{abstract}
This paper derives a \emph{distributed} Kalman filter to estimate a sparsely connected, large-scale, $n-$dimensional, dynamical system monitored by a network of~$N$ sensors. Local Kalman filters are implemented on the ($n_l-$dimensional, where $n_l\ll n$) sub-systems that are obtained after spatially decomposing the large-scale system. The resulting sub-systems overlap, which along with an assimilation procedure on the local Kalman filters, preserve an $L$th order Gauss-Markovian structure of the centralized error processes. The information loss due to the $L$th order Gauss-Markovian approximation is controllable as it can be characterized by a divergence that decreases as $L\uparrow$. The order of the approximation, $L$, leads to a lower bound on the dimension of the sub-systems, hence, providing a criterion for sub-system selection. The assimilation procedure is carried out on the local error covariances with a distributed iterate collapse inversion (DICI) algorithm that we introduce. The DICI algorithm computes the (approximated) centralized Riccati and Lyapunov equations iteratively with only local communication and low-order computation. We fuse the observations that are common among the local Kalman filters using bipartite fusion graphs and consensus averaging algorithms. The proposed algorithm achieves full distribution of the Kalman filter that is coherent with the centralized Kalman filter with an $L$th order Gaussian-Markovian structure on the centralized error processes. Nowhere storage, communication, or computation of~$n-$dimensional vectors and matrices is needed; only~$n_l \ll n$ dimensional vectors and matrices are communicated or used in the computation at the sensors. \end{abstract}

\newpage
\section{Introduction}
Centralized implementation of the Kalman filter~\cite{kalman:60,kalman:61}, although possibly optimal, does not provide robustness and scalability when it comes to complex large-scale dynamical systems with their measurements distributed on a large geographical region. The reasons are twofold:~\begin{inparaenum}[(i)]\item the large-scale systems are very high-dimensional, and thus extensive computations are required to implement the centralized procedure;~and~\item the span of the geographical region, over which the large-scale system is deployed or the physical phenomenon is observed, poses a large communication burden and thus, among other problems, adds latency to the estimation mechanism\end{inparaenum}. To remove the difficulties posed by the centralized procedure, we propose a distributed estimation algorithm with a low order Kalman filter at each sensor. To account for the processing, communication, and limited resources at the sensors, the local Kalman filters involve computations and communications with local quantities only, i.e., vectors and matrices of low dimensions, $n_l\ll n$, where $n$ is the dimension of the state vector---no sensor computes, communicates, or stores any $n-$dimensional quantity.

Much of the existing research on distributed Kalman filters focuses on sensor networks monitoring low dimensional systems, e.g., when multiple sensors mounted on a small number of robot platforms are used for target tracking~\cite{raowhyte:91,salig:06,murray:04}. This scenario addresses the problem of how to efficiently incorporate the distributed observations, which is also referred to in the literature as `data fusion,' see also~\cite{hashem:88}. Data fusion for Kalman filters over arbitrary communication networks is discussed in~\cite{olfati:05}, using consensus protocols in~\cite{olfati2:05}. References~\cite{sinopoli:04,salig:06} incorporate packet losses, intermittent observations, and communication delays in the data fusion process. Although these solutions work well for low-dimensional dynamical systems, e.g., target tracking applications, they implement a replication of the~$n-$dimensional Kalman filter at each sensor, hence, communicating and inverting~$n \times n$ matrices locally, which, in general, is an $O(n^3)$ operation. In contrast, in the problems we consider, the state dimension,~$n$, is very large, for example, in the range of~$10^2$ to~$10^9$, and so the~$n$th order replication of the global dynamics in the local Kalman filters is either not practical or not possible.

Kalman filters with reduced order models have been studied, in e.g.,~\cite{bergwhyte:90,Mutambara_book} to address the computation burden posed by implementing $n$th order models. In these works, the reduced models are decoupled, which is sub-optimal as important coupling among the system variables is ignored. Furthermore, the network topology is either fully connected~\cite{bergwhyte:90}, or is close to fully connected~\cite{Mutambara_book}, requiring long distance communication that is expensive. We are motivated by problems where the large-scale systems, although sparse, cannot be decoupled, and where, due to the sensor constraints, the communication and computation should both be local. Multi-level systems theory to derive a coordination algorithm among local Kalman filters is discussed in \cite{noton:71}.

We present a distributed Kalman filter that addresses {\it both} the computation and communication challenges posed by complex large-scale dynamical systems, while preserving its coupled structure; in particular, nowhere in our distributed Kalman filter, do we require storage, communication, or computation of~$n-$dimensional quantities. We briefly explain the key steps and approximations in our solution.

{\bf Spatial Decomposition of Complex Large-Scale Systems:} To distribute the Kalman filter, we provide a spatial decomposition of the complex large-scale dynamical system (of dimension $n$) that we refer to as {\it the overall system} into several, possibly many, local coupled dynamical systems (of dimension $n_l$, such that $n_l\ll n$) that we refer to as {\it sub-systems} in the following. The large-scale systems, we consider, are sparse and localized. Physical systems with such characteristics are described in section~\ref{glob_model}. We exploit the underlying distributed and sparse structure of system dynamics that results from a spatio-temporal discretization of random fields. Dynamical systems with such structure are then spatially decomposed into sub-systems. The sub-systems overlap and thus the resulting local Kalman filters also overlap. This overlap along with an assimilation procedure on the local error covariances (of the local filters) preserve the structure of the centralized (approximated) error covariances. We preserve the coupling by applying the coupled state variables as inputs to the sub-systems. In contrast, an estimation scheme using local observers is addressed in~\cite{sys_dig_book}, where the coupled states are also applied as inputs to the sub-systems, but, the error covariances are not assimilated and remain decoupled. Such scheme loses its coherence with the centralized procedure, as under arbitrary coupling, no structure of the centralized error covariance can be retained by the local filters.

{\bf Overlapping Dynamics at the Sub-systems: Bipartite Fusion Graphs:} The sub-systems that we extract from the overall system overlap. This makes several state variables to be observed at different sub-systems. To fuse this shared information, we implement a fusion algorithm using bipartite fusion graphs, which we introduced in~\cite{usman_sspw:07}, and local average consensus algorithms~\cite{boyd:04,olfati2:05}. The interactions required by the fusion procedure are constrained to a small neighborhood (this may require multi-hop communication in the small neighborhood). The multi-hop communication requirements, in our case, arise because the dynamical field is distributed among local sub-systems and only local processing is carried out throughout the development. In contrast, as noted earlier, a distributed Kalman filter with single hop communication scheme is presented in \cite{olfati:05}, but, we reemphasize the fact the this single-hop communications is a result of replicating $n$th order Kalman filters at each sensor.

{\bf Assimilation of the Local Error Covariances---Distributed Iterate-Collapse Inversion (DICI) Algorithm:} A key issue in distributing the Kalman filter is to maintain a reasonable coherence with the centralized estimation; this is to say that the local error covariances should approximate the centralized error covariance in a well-defined manner. If the local error covariances evolve independently at each sub-system they may lose any coherence with the centralized error covariance under arbitrary coupling. To address this issue, we employ a {\it cooperative assimilation procedure} among the local error covariances that preserves a Gauss-Markovian structure in the centralized error processes. This is equivalent to approximating\footnote{A measure on the loss of optimality is provided in section~\ref{cLBIF} and it will be shown that this measure serves as a criterion for choosing the dimensions of the sub-systems.} the inverse of the centralized error covariances (the information matrices) to be~$L-$banded\footnote{We refer to a matrix as an~$L$-banded matrix ($L\geq0$), if the elements outside the band defined by the~$L$th upper and~$L$th~lower~diagonal~are~0.}, as shown in~\cite{balram:93}. The assimilation procedure is carried out with a distributed iterate-collapse inversion (DICI, pronounced die-see) algorithm, briefly introduced in~\cite{usman_icassp:07}, that has an iterate step and a collapse step.

We use the Information filter, \cite{Moore_book, Mutambara_book}, format of the Kalman filter where the information matrices (inverse of the error covariances) are iterated at each time step. We introduce the DICI algorithm that provides an iterative matrix inversion procedure to obtain the local error covariances from the local information matrices. The inverses of the local information matrices are assimilated among the sub-systems such that the local error covariances, obtained after the assimilation, preserve the Gauss-Markovian structure of the centralized error covariances. Iterative matrix inversion can also be carried out using the distributed Jacobi algorithm~\cite{tsit_book} where the computational complexity scales linearly with the dimension, $n$, of the overall system. In contrast, the computational complexity of the DICI algorithms is independent of $n$. In addition, the error process of the DICI algorithm is bounded above by the error process of the distributed Jacobi algorithm. We show the convergence of the iterate step of the DICI algorithm analytically and resort to numerical simulations to show the convergence of its collapse step.

In summary, the spatial decomposition of the complex large-scale systems, fusion algorithms for fusing observations, the DICI algorithm to assimilate the local error covariance combine to give a robust, scalable, and distributed implementation of the Kalman filter.

We describe the rest of the paper. Section~\ref{bkg} covers the discrete-time models, centralized Information filters, and centralized~$L-$banded Information filters. Section~\ref{mod_dist} covers the model distribution step. We introduce the local Information filters in section~\ref{LIF:sub_mod} along with the necessary notation. Section~\ref{LIF:orm} gives the observation fusion step of the local Information filters, and section~\ref{dist_jac} presents the distributed iterate collapse inversion (DICI) algorithm. The filter step of the local Information filters is provided in section~\ref{LIF:ic_lfs}, and the prediction step of the local Information filters is provided in section~\ref{LIF:lps}. We conclude the paper with results in section~\ref{res} and conclusions in section~\ref{conc}. Appendix~\ref{LBth} discusses the~$L-$banded inversion theorem,~\cite{kavcic:00}.

\section{Background}\label{bkg}
\PARstart{I}n this section, we motivate the type of applications and large-scale dynamical systems of interest to us. The context is that of a time-varying random field governed by partial differential equations (PDEs); these systems can also be generalized to arbitrary dynamical systems belonging to a particular structural class, as we elaborate in Subsection~\ref{glob_model}. To fix notation, we then present the centralized version of the Information filter.

\subsection{Global Model}
\label{glob_model}
Many physical phenomenon~\cite{nash:78,staelin:78,brammer:79,njoku:99,buehner:03,inui:03}, e.g., ocean/wind circulation and heat/wave equations, can be broadly characterized by a PDE of the Navier-Stokes type. These are highly non-linear and different regimens arise from different assumptions. Non-linear approximation models are commonly used for prediction (predictive models). For data assimilation, i.e., combining models with measured data, e.g., satellite altimetry data in ocean models, it is unfeasible to use non-linear models; rather linearized approximations (dynamical linearization) are employed. Here, our goal is to motivate how discrete linear models occur that exhibit a sparse and localized structure that we use to distribute the model in Section~\ref{mod_dist}. Hence, we take a very simplistic example and consider elliptical operators, given by, \begin{equation}\label{ellip}\mathcal{L}=\alpha\dfrac{\partial^2}{\partial \rho_x^2}+\beta \dfrac{\partial^2}{\partial \rho_y^2},\end{equation} where $\rho_x$ and $\rho_y$ represent the horizontal and vertical dimensions, respectively, and $\alpha,\beta$ are constants pertinent to the specific application. The spatial model for the continuous-time physical phenomenon (e.g., heat, wave, or, wind), $x_t$, with a continuous-time random noise term, $u_t$, can now be written as, \begin{equation}\label{gen_mod}\mathcal{L} x_t = u_t,\end{equation} which reduces to a Laplacian equation, $\Delta x_t=0~(\Delta={\partial^2}/{\partial \rho_x^2}+{\partial^2}/{\partial \rho_y^2})$, or to a Poisson equation, $-\Delta x_t=u_t$, for appropriate choices of the constants. We spatially discretize the general model in~\eqref{gen_mod} on an $M\times J$ uniform mesh grid, where the standard 2nd order difference equation approximation to the 2nd order derivative is \begin{eqnarray}\dfrac{\partial^2 x_t}{\partial \rho_x^2}\sim x_{i+1,j} - 2x_{i,j} + x_{i-1,j},\qquad\dfrac{\partial^2 x_t}{\partial \rho_y^2} \sim x_{i,j+1} -2x_{i,j} + x_{i,j-1},\end{eqnarray} where $x_{ij}$ is the value of the random field, $x_t$, at the $ij$-th location in the $M\times J$ grid. With this approximation, the Laplace equation (say), leads to the individual equations of the form \begin{equation}\label{ind_eq}x_{i,j}-\frac{1}{4}\left(x_{i,j-1}+x_{i,j+1}+x_{i-1,j}+x_{i+1,j}\right)=0.\end{equation} Let $\mathbf{I}$ be an identity matrix and let $\mathbf{A}$ be a tridiagonal matrix with zeros on the main diagonal and ones on the upper and lower diagonals, then \eqref{ind_eq}, $\forall~i,j \in M\times J$ (with appropriate constants for modeling the elliptical operator in \eqref{gen_mod}), can be collected in the linear system of equations and \eqref{gen_mod} can be spatially discretized in the form\begin{equation}\mathbf{F}_c\mathbf{x}_t=\mathbf{b}_c+\mathbf{G}_c \mathbf{u}_t,\end{equation} where the vector $\mathbf{b}_c$ collects the boundary conditions, the term $\mathbf{G}_c\mathbf{u}_t$ controls where the random noise is added and \begin{equation}\mathbf{x}_t=\left[\left(\mathbf{x}_t^1\right)^T,\ldots,\left(\mathbf{x}_t^M\right)^T\right]^T\end{equation} and $\mathbf{x}_t^i = [x_{i1},\ldots,x_{iJ}]^T$. The matrix $\mathbf{F}_c$ is given by \begin{eqnarray}\label{PDE_F}\mathbf{F}_c &=& \left[\begin{array}{ccc}
\mathbf{B}&\mathbf{C}&\\
\ddots&\ddots&\ddots\\
&\mathbf{C}&\mathbf{B}\end{array}
\right]=\mathbf{I}\otimes\mathbf{B} + \mathbf{A}\otimes \mathbf{C},\end{eqnarray} where $\mathbf{B}=\mu\mathbf{I} + \beta_h\mathbf{A},$ and $\mathbf{C} = \beta_v\mathbf{I},$ the constants~$\mu,\beta_h,\beta_v$ are in terms of $\alpha,\beta$ in \eqref{ellip}, and $\otimes$ is the Kronecker product~\cite{kron_book}.

%\begin{eqnarray}\label{PDE_F}\mathbf{F}_c &=& \left[\begin{array}{ccccc}
%\mathbf{B}&\mathbf{C}&&&\\
%\mathbf{C}&\mathbf{B}&\mathbf{C}&&\\
%&\ddots&\ddots&\ddots&\\
%&&\mathbf{C}&\mathbf{B}&\mathbf{C}\\
%&&&\mathbf{C}&\mathbf{B}\end{array}
%\right],\nonumber\\&=&\mathbf{I}\otimes\mathbf{B} + \mathbf{A}\otimes \mathbf{C},\end{eqnarray}
%
If we further include time dependence, e.g., by discretizing the diffusion equation, we get \begin{equation}\label{c2dmod}\mathbf{\dot{x}}_t=\mathbf{F}_c\mathbf{x}_t-\mathbf{b}_c-\mathbf{G}_c\mathbf{u}_t,\end{equation} where $\mathbf{\dot{x}}_t=d\mathbf{x}_t/dt$. We again employ the standard first difference approximation of the first derivative with $\triangle_T $ being the sampling interval and $k$ being the discrete-time index, and, write the above equation as \begin{equation}\mathbf{x}_{k+1} = \left(\mathbf{I}+\triangle_T \mathbf{F}_c \right) \mathbf{x}_k - \triangle_T \mathbf{b}_c - \triangle_T \mathbf{G}_c\mathbf{u}_k, \qquad\qquad k\geq 0 \end{equation} which can be generalized as the following discrete-time model (where for simplicity of the presentation and without loss of generality we drop the term $\mathbf{b}_c$ in the sequel) \begin{equation}\label{glob_dyn}
\mathbf{x}_{k+1} =\mathbf{Fx}_{k}+\mathbf{Gu}_{k}.\end{equation}In the above model,~$\mathbf{x}_k \in \mathbb{R}^n$ is the state vector,~$\mathbf{x}_0 \in \mathbb{R}^n$ are the initial state conditions,~$\mathbf{F} \in \mathbb{R}^{n \times n}$ is the model matrix,~$\mathbf{u}_k \in \mathbb{R}^{j\prime}$ is the state noise vector and~$\mathbf{G} \in \mathbb{R}^{n \times j\prime}$ is the state noise matrix.

Here, we note that the model matrix, $\mathbf{F}$, is is perfectly banded in case of PDEs as in \eqref{PDE_F}. We can relax this to sparse and localized, i.e., the coupling among the states decays with distance (in an appropriate measure). Besides random fields, the discrete-space-time model \eqref{glob_dyn} with sparse and localized structure also occurs, e.g., in image processing, the dynamics at a pixel depends on neighboring pixel values~\cite{katsaggelos:95},~\cite{fieguth:05}; in power systems, the power grid models, under certain assumptions, exhibit banded structures,~\cite{hill:81,ilic:05,usman_pse:07}. Systems that are sparse but not localized can be converted to sparse and localized by using matrix bandwidth reduction algorithms~\cite{mckee:69}. We further assume that the overall system is coupled, irreducible, and globally observable.

Let the system described in~\eqref{glob_dyn} be monitored by a network of~$N$ sensors. Observations at sensor~$l$~and~time~$k$~are \begin{equation}\label{loc_obs}\mathbf{y}_{k}^{(l)}=\mathbf{H}_{l}\mathbf{x}_{k}+\mathbf{w}%
_{k}^{(l)},\end{equation} where~${\mathbf{H}_{l}} \in \mathbb{R}^{p_l\times n}$ is the local observation matrix for sensor~$l$,~$p_l$ is the number of simultaneous observations made by sensor~$l$ at time $k$, and~$\mathbf{w}_k^{(l)} \in \mathbb{R}^{p_l}$ is the local observation noise. In the context of the systems we are interested in, it is natural to assume that the observations are localized. These local observations at sensor $l$ may be, e.g., the temperature or height at location $l$ or an average of the temperatures or heights at $l$ and neighboring locations.

We stack the observations at all~$N$ sensors in the sensor network to get the global observation model as follows. Let~$p$ be the total number of observations at all the sensors. Let the global observation vector,~$\mathbf{y}_k\in\mathbb{R}^{p}$, the global observation matrix, $\mathbf{H}\in\mathbb{R}^{p\times n}$, and the global observation noise vector,~$\mathbf{w}_k\in\mathbb{R}^{p}$, be \begin{align}\mathbf{y}_k&=\left[\begin{array}{c}\mathbf{y}_k^{(1)}\\\vdots\\\mathbf{y}_k^{(N)}\end{array}\right],& \label{H} \mathbf{H}&=\left[\begin{array}{c}\mathbf{H}_1\\\vdots\\\mathbf{H}_N\end{array}\right],&\mathbf{w}_k&=\left[\begin{array}{c}\mathbf{w}_k^{(1)}\\\vdots\\\mathbf{w}_k^{(N)}
\end{array}\right].&\end{align}

Then the global observation model is given by
\begin{equation}\label{glob_obs}
\mathbf{y}_{k}=\mathbf{Hx}_{k}+\mathbf{w}_{k}.
\end{equation}

We adopt standard assumptions on the statistical characteristics of the noise. The state noise sequence, $\{\mathbf{u}_k\}_{k\geq0}$, the observation noise sequence,~$\{\mathbf{w}_k\}_{k\geq0}$, and the initial conditions, $\mathbf{x}_0$, are independent,~Gaussian,~zero-mean,~with \begin{equation}\mathbb{E}[\mathbf{u}_\iota\mathbf{u}_\tau^H] = \mathbf{Q}\delta _{\iota\tau}\mbox{ and }\mathbb{E}[\mathbf{w}_\iota\mathbf{w}_\tau^H] = \mathbf{R}\delta _{\iota\tau},\mbox{  and   }\label{Sigma}\mathbb{E}[\mathbf{x}_0\mathbf{x}_0^H]=\mathbf{S}_0,\end{equation} where the superscript $H$ denotes the Hermitian, the Kronecker delta~$\delta_{\iota\tau}=1$, if and only if~$\iota=\tau$, and zero otherwise. Since the observation noise at different sensors is independent, we can partition the global observation noise covariance matrix,~$\mathbf{R}$, with~$\mathbf{R}_l\in \mathbb{R}^{p_l\times p_l}$ being the local observation noise covariance matrix at sensor~$l$, as \begin{equation}\label{R}\mathbf{R}=\mbox{blockdiag}[\mathbf{R}_1,\ldots,\mathbf{R}_N].\end{equation}

For the rest of the presentation, we consider time-invariant models, specifically, the matrices, $\mathbf{F,G,H,Q,R}$, are time-invariant. The discussion, however, is not limited to either zero-mean initial condition or time-invariant models and generalizations to the time-variant models will be added as we proceed.

\subsection{Centralized Information Filter}
\label{cIF}
Let~$\mathbf{\breve{S}}_{k|k}$ and~$\mathbf{\breve{S}}_{k|k-1}$ be (filtered and prediction) error covariances, and their inverses be the information matrices,~$\mathbf{\breve{Z}}_{k|k}$ and~$\mathbf{\breve{Z}}_{k|k-1}$. Let~$\mathbf{\widehat{x}}_{k|k}$ and~$\mathbf{\widehat{x}}_{k|k-1}$ be the filtered estimate and the predicted estimate of the state vector,~$\mathbf{x}_k$, respectively. We have the following relations. \begin{eqnarray}\label{Z2S}\mathbf{\breve{S}}_{k|k}&=&\mathbf{\breve{Z}}_{k|k}^{-1}\\\label{Z2Pi}
\mathbf{\breve{S}}_{k|k-1}&=&\mathbf{\breve{Z}}_{k|k-1}^{-1}\\\label{zp2xp}
\mathbf{\widehat{z}}_{k|k-1}&=&\mathbf{\breve{Z}}_{k|k-1}\mathbf{\widehat{x}}_{k|k-1}\\ \label{z2x}\mathbf{\widehat{z}}_{k|k}&=&\mathbf{\breve{Z}}_{k|k}\mathbf{\widehat{x}}_{k|k}
\end{eqnarray} Define the~$n-$dimensional global observation variables as%\footnote{The information variable, $\mathbf{\mathcal{I}}$, is time dependent, if the global observation matrix, $\mathbf{H}$, or the observation noise covariance matrix, $\mathbf{R}$, were time-dependent. In this case, it is given by $$\mathcal{I}_{k}=\mathbf{H}_k^{T}\mathbf{R}_k^{-1}\mathbf{H}_k.$$}
\begin{eqnarray}\label{ik}\mathbf{i}_{k}&=&\mathbf{H}^{T}\mathbf{R}^{-1}\mathbf{y}_{k},\\\label{Ik}
\mathbf{\mathcal{I}}&=&\mathbf{H}^{T}\mathbf{R}^{-1}\mathbf{H},\end{eqnarray} and the~$n-$dimensional local observation variables at sensor~$l$ as \begin{eqnarray}\label{ilk}\mathbf{i}_{l,k}&=&\mathbf{H}_l^{T}\mathbf{R}_l^{-1}\mathbf{y}_{k}^{(l)},\\\label{Ilk}
\mathbf{\mathcal{I}}_{l}&=&\mathbf{H}_l^{T}\mathbf{R}_l^{-1}\mathbf{H}_l.\end{eqnarray} When the observations are distributed among the sensors, see~\eqref{loc_obs}, the CIF can be implemented by collecting all the sensor observations at a central location; or with observation fusion by realizing that the global observation variables in~\eqref{ik}$-$\eqref{Ik}, can be written as, see~\cite{raowhyte:91,Mutambara_book,olfati:05},\begin{eqnarray}\label{ikf}
\mathbf{i}_{k} &\mathbf{=}&\mathbf{H}^{T}\mathbf{R}^{-1}\mathbf{y}_{k} \nonumber \\
&=&\mathbf{H}_{1}^{T}\mathbf{R}_{1}^{-1}\mathbf{y}_{k}^{(1)}+\cdots +%
\mathbf{H}_{N}^{T}\mathbf{R}_{N}^{-1}\mathbf{y}_{k}^{(N)} \nonumber\\ &=& \sum_{l=1}^{N} \mathbf{i}_{l,k}.
\end{eqnarray} Similarly, \begin{equation}\label{Ikf}
\mathbf{\mathcal{I}}=\sum_{l=1}^{N}\mathbf{\mathcal{I}}_{l}.
\end{equation} The {\it filter step} of the CIF is
\begin{subequations}\label{glob_filt_step}
\begin{eqnarray}\label{glob_filt_step_pt1}
\mathbf{\breve{Z}}_{k|k} &=& \mathbf{\breve{Z}}_{k|k-1}+\sum\limits_{l=1}^{N}\mathbf{\mathcal{I}}_{l},\qquad\qquad k\geq 0\\\label{glob_filt_step_pt2}
\mathbf{\widehat{z}}_{k|k}&=&\mathbf{\widehat{z}}_{k|k-1}+\sum_{l=1}^{N}\mathbf{i}_{l,k},\qquad\qquad k\geq 0.
\end{eqnarray}
\end{subequations} The {\it prediction step} of the CIF is
\begin{subequations}\label{glob_pred_step}
\begin{eqnarray}\label{g_pred_pt1}
\mathbf{\breve{Z}}_{k|k-1} &=& \mathbf{\breve{S}}_{k|k-1}^{-1} = \mathbf{(F}\mathbf{\breve{Z}}_{k-1|k-1}
^{-1}\mathbf{F}^{T}\mathbf{+GQG}^{T}\mathbf{)}^{-1},\qquad\qquad k\geq1,~\mathbf{\breve{Z}}_{0|-1} = \mathbf{\breve{S}}_0^{-1},\\\label{g_pred_pt2}
\mathbf{\widehat{z}}_{k|k-1} &=&\mathbf{\breve{Z}}_{k|k-1}\left(\mathbf{F\breve{Z}}_{k-1|k-1}^{-1}\mathbf{\widehat{z}}_{k-1|k-1}\right),\qquad\qquad\qquad~~~~k \geq1,~\mathbf{\widehat{z}}_{0|-1}=\mathbf{0}.
\end{eqnarray}
\end{subequations}

%The CIF needs: (i) the knowledge of all the observations,~$\mathbf{y}_k$, at a central location to compute~\eqref{ikf}, a non-trivial communication task when the number of sensors,~$N$, is large; and (ii) global filter computations, e.g.,~\eqref{glob_pred_step}, an infeasible challenge when the number of states,~$n$, is very large. Further, the CIF has the disadvantages of large latency and a single point of failure.
%
\subsection{Centralized~$L-$Banded Information filters}\label{cLBIF}
To avoid the~$O(n^3)$ computations of the global quantities in~\eqref{glob_pred_step}, e.g., the inversion,~$\mathbf{\breve{Z}}_{k-1|k-1}^{-1}$, we approximate the information matrices,~$\mathbf{\breve{Z}}_{k|k}$ and~$\mathbf{\breve{Z}}_{k|k-1}$, to be~$L-$banded matrices,~$\mathbf{Z}_{k|k}$ and~$\mathbf{Z}_{k|k-1}$. We refer to the CIF with this approximation as the centralized $L-$banded Information filter (CLBIF). This approach is studied in~\cite{chin:92}, where the information loss between $\mathbf{\breve{Z}}$ and $\mathbf{Z}$, is given by the divergence (see~\cite{chin:92})\begin{eqnarray} \label{divL}\mbox{Divergence}(\mathbf{\breve{Z},Z})&=&\frac{1}{2}\left|\left| \mathbf{\breve{Z}}^{-\frac{T}{2}}\left(\mathbf{Z}-\mathbf{\breve{Z}}\right)\mathbf{{Z}}^{-\frac{1}{2}}\right|\right|^2_F \leq\frac{1}{2}\left(\sum_i\lambda_{i(\mathbf{Z})}^{-\frac{1}{2}}\right)^2\left(\sum_i\lambda_{i(\mathbf{\breve{Z}})}^{-\frac{1}{2}} \right)^2\left|\left|\mathbf{Z}-\mathbf{\breve{Z}}\right|\right|^2_F,\end{eqnarray}where $||\cdot||_F$ is the Frobenius norm and $\lambda_{i(\mathbf{Z})}$ is the $i$th eigenvalue of the matrix~$\mathbf{Z}$.

This approximation on the information matrices is equivalent to approximating the Gaussian error processes, \begin{eqnarray}\mathbf{\epsilon}_{k|k}&=&\mathbf{x}_k-\mathbf{\widehat{x}}_{k|k},\qquad\mathbf{\epsilon}_{k|k-1}=\mathbf{x}_k-\mathbf{\widehat{x}}_{k|k-1},\end{eqnarray} to be Gauss-Markovian of~$L$th order~\cite{balram:93}. Reference~\cite{kavcic:00} presents an algorithm to derive the approximation that is optimal in Kullback-Leibler or maximum entropy sense in the class of all $L-$banded matrices approximating the inverse of the error covariance matrix. In the sequel, we assume this optimal $L-$banded approximation.

The CLBIF (with the~$L-$banded information matrices,~$\mathbf{Z}_{k|k}$ and~$\mathbf{Z}_{k|k-1}$) is given by the {\it filter step} in \eqref{glob_filt_step_pt1}$-$\eqref{glob_filt_step_pt2} and the {\it prediction step} in~\eqref{g_pred_pt1}$-$\eqref{g_pred_pt2}, where the optimal information matrices,~$\mathbf{\breve{Z}}_{k|k}$ and~$\mathbf{\breve{Z}}_{k|k-1}$, are replaced by their $L-$banded approximations. The algorithms in \cite{kavcic:00,asif:05} reduce the computational complexity of the CLBIF to $O(n^2)$ but the resulting algorithm is still centralized and deals with the~$n-$dimensional state. To distribute the CLBIF, we start by distributing the global model~\eqref{glob_dyn}$-$\eqref{glob_obs} in the following section.

\section{Spatial Decomposition of Complex Large-Scale Systems}\label{mod_dist}
Instead of implementing CLBIF based on the global model, we implement local Information filters at the sub-systems obtained by spatially decomposing the overall system. Subsection~\ref{red_mod_sen} deals with this decomposition by exploiting the sparse and localized structure of the model matrix,~$\mathbf{F}$. In the following, we devise a sub-system at each sensor monitoring the system. This can be applied to general model matrices, $\mathbf{F}$, but is only practical if $\mathbf{F}$ is sparse and localized as we explain below.

\subsection{Reduced Models at Each Sensor}
\label{red_mod_sen}
This subsection shows how to distribute the global model~\eqref{glob_dyn} and~\eqref{glob_obs}, in order to get the reduced order sub-systems. We illustrate the procedure with a simple example that reflects our assumptions on the dynamical system structure. Consider a five dimensional system with the global dynamical model \begin{eqnarray}\label{ex_sys1}\mathbf{x}_{k+1} &=&\left[
\begin{array}{ccccc}
f_{11} & f_{12} & 0 & 0 & 0 \\
f_{21} & f_{22} & 0 & f_{24} & 0 \\
f_{31} & 0 & f_{33} & 0 & 0 \\
0 & 0 & f_{43} & 0 & f_{45} \\
0 & 0 & 0 & f_{54} & f_{55}%
\end{array}%
\right]\mathbf{x}_k + \left[
\begin{array}{cc}
0 & 0\\
0 & 0\\
0 & g_{32} \\
0 & 0 \\
g_{51} & 0%
\end{array}%
\right]\mathbf{u}_k\\
&=&\mathbf{Fx}_k+\mathbf{Gu}_k \nonumber.\end{eqnarray} The system has two external noise sources~$\mathbf{u}_k=[u_{1k},u_{2k}]^T$. We monitor this system with~$N= 3$ sensors, having scalar observations,~$y_k^{(l)}$, at each sensor~$l$. The global observation vector,~$\mathbf{y}_k$, stacks the local observations,~$y_k^{(l)}$,~and~is\begin{eqnarray}\label{ex_sys2}
\mathbf{y}_k=\left[
\begin{array}{c}
y^{(1)}_k \\
y^{(2)}_k \\
y^{(3)}_k%
\end{array}%
\right]&=&
\left[
\begin{array}{ccccc}
1 & 1 & 1 & 0 & 0 \\
0 & 1 & 1 & 1 & 0 \\
0 & 0 & 0 & 1 & 1%
\end{array}%
\right]\mathbf{x}_k + \left[
\begin{array}{c}
w^{(1)}_k \\
w^{(2)}_k \\
w^{(3)}_k%
\end{array}%
\right]\\
&=& \mathbf{Hx}_k + \mathbf{w}_k, \nonumber
\end{eqnarray} where~$\mathbf{H}=[\mathbf{H}_1^T~\mathbf{H}_2^T~\mathbf{H}_3^T]^T$. We distribute global model of equations~\eqref{ex_sys1} and~\eqref{ex_sys2} in the following Subsections.

\subsubsection{Graphical Representation using System Digraphs}\label{sysD}
\begin{figure}
\centering
\subfigure[]
{
    \label{sys_d}
    \includegraphics[width=1.5in]{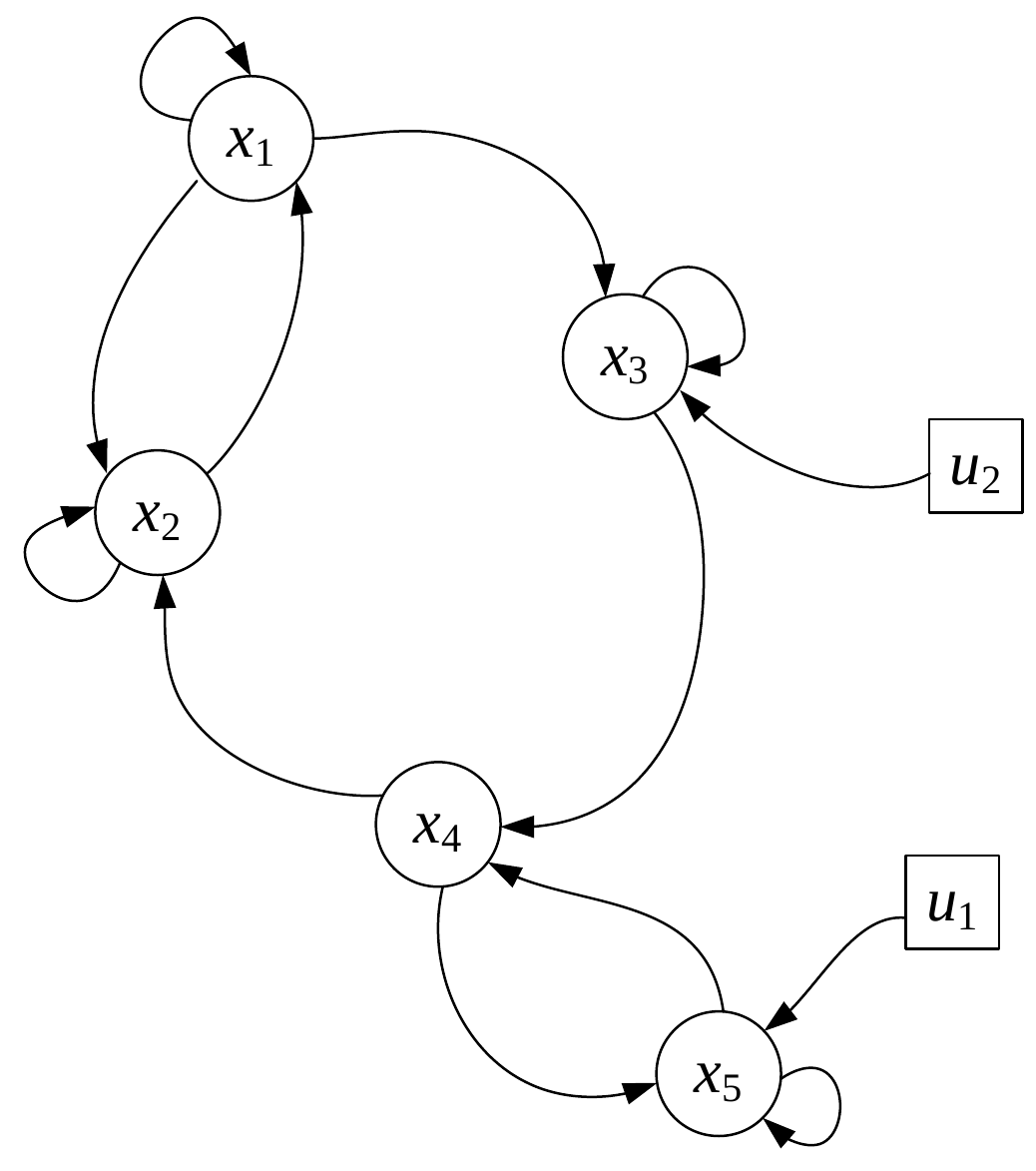}
}
\hspace{3cm}
\subfigure[]
{
    \label{cutsets_fig.pdf}
    \includegraphics[width=1.5in]{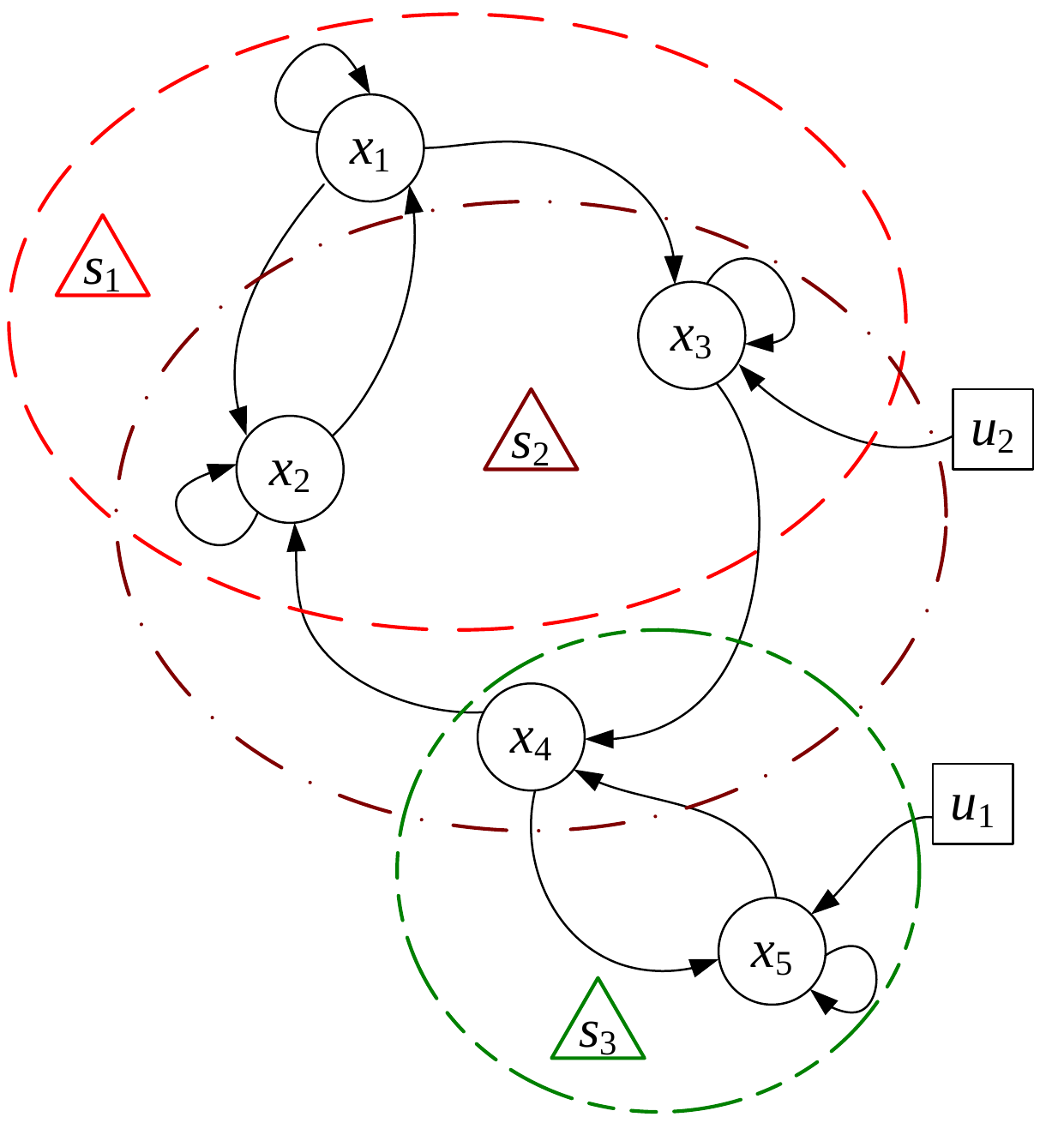}
}
\caption{System Digraph and cut-point sets: (a) Digraph representation of the~$5$ dimensional system,~\eqref{ex_sys1}$-$\eqref{ex_sys2}. The circles represent the states,~$\mathbf{x}$, and the squares represent the input noise sources,~$\mathbf{u}$. (b) The cut-point sets associated to the~$3$ sensors ($\triangle$) are shown by the dashed circles.}
\label{sys_cut_fig} % caption for the whole figure
\end{figure}

A system digraph visualizes the dynamical interdependence of the system. A system digraph,~\cite{sys_dig_book}, $\mathcal{J} = [V, E]$, is a directed graphical representation of the system, where~$V=X\cup U$ is the vertex set consisting of the states,~$X=\{x_i\}_{i=1,\ldots,n}$, and the noise inputs,~$U=\{u_i\}_{i=1,\ldots,j}$. The interconnection matrix, $E$, is the binary representation (having a~$1$ for each non-zero entry) of the model matrix,~$\mathbf{F}$, and the state noise matrix,~$\mathbf{G}$, concatenated together. The interconnection matrix,~$E$, for the system in~\eqref{ex_sys1} is, \begin{equation}E=\left[
\begin{array}{ccccccc}
1 & 1 & 0 & 0 & 0 & 0 & 0 \\
1 & 1 & 0 & 1 & 0 & 0 & 0 \\
1 & 0 & 1 & 0 & 0 & 0 & 1 \\
0 & 0 & 1 & 0 & 1 & 0 & 0 \\
0 & 0 & 0 & 1 & 1 & 1 & 0%
\end{array}%
\right].\end{equation} The system digraph is shown in Figure~\ref{sys_d}.

\subsubsection{Reduced Models from the Cut-point Sets}\label{cutP}
We have~$N=3$ sensors monitoring the system through the observation model~\eqref{ex_sys2}. We associate to each sensor~$l$ a {\it cut-point set}, $V^{(l)}$, where~$V^{(l)}\subseteq X$. If a sensor observes a linear combination of several states, its cut-point set,~$V^{(l)}$, includes all these states; see~\cite{haotian:05} for the definition of cut-point sets, and algorithms to find all cut-point sets and a minimal cut-point set, if it exists. The cut-point sets select the local states involved in the local dynamics at each sensor. From~\eqref{ex_sys2}, the cut-point sets\footnote{For simplicity of the presentation, we chose here that each state variable is observed by at least one sensor. If this is not true, we can easily account for this by extending the cut-point sets,~$V^{(l)}$, to~$\overline{V}^{(l)}$, such that \begin{equation}\bigcup_{l=1}^N \overline{V}^{(l)} = X.\label{cutext}\end{equation}} are shown in Figure~\ref{cutsets_fig} where we have the following cut-point set, e.g., at sensor $1$,\begin{eqnarray}\label{V1}V^{(1)} &=& \left\{x_1,x_2,x_3\right\}.\end{eqnarray} \begin{figure}
\centering
\includegraphics[width=1.5in]{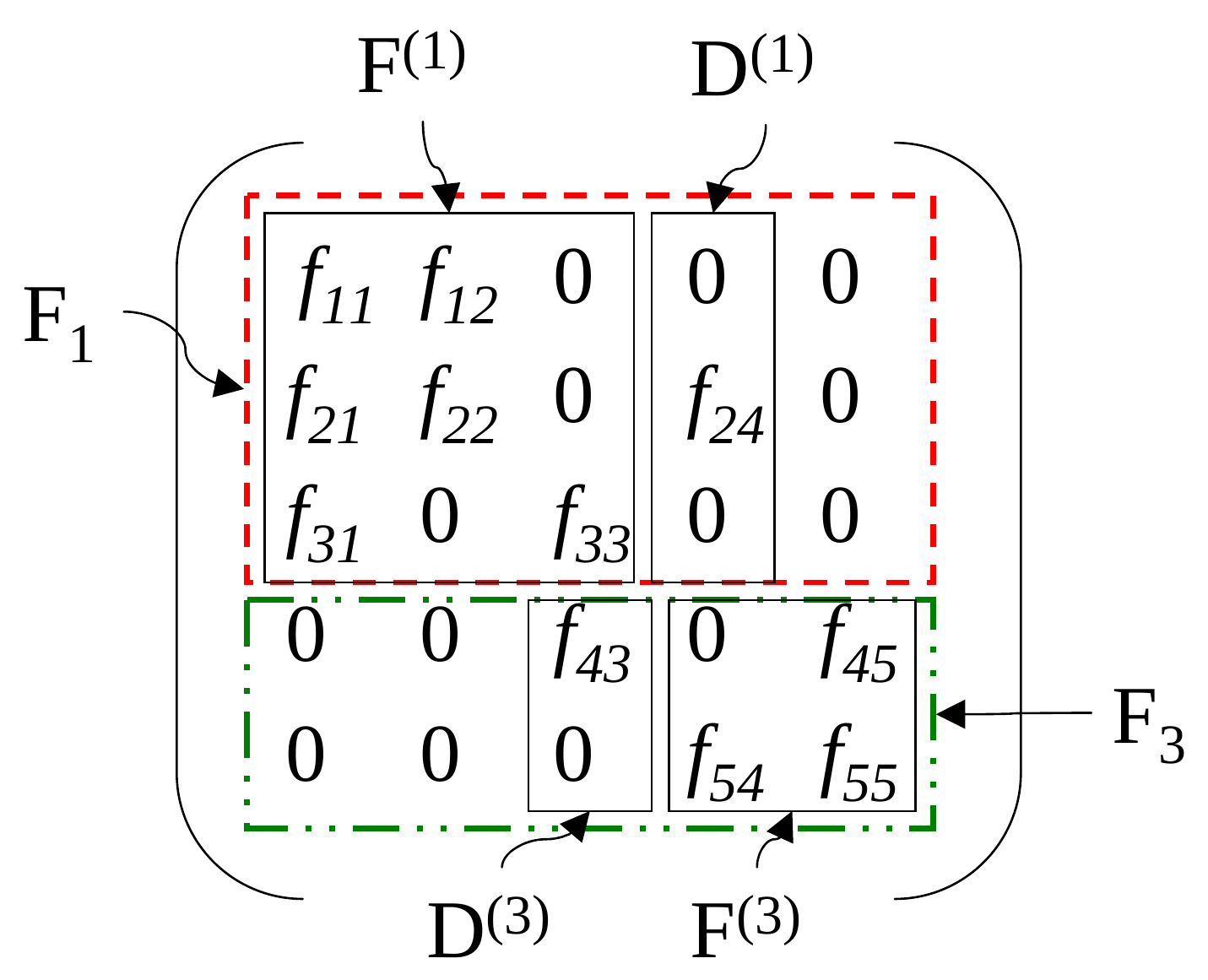}
\caption{Partitioning of the global model matrix,~$\mathbf{F}$, into local model matrices,~$\mathbf{F}^{(l)}$, and the local internal input matrices, $\mathbf{D}^{(l)}$, shown for sensor~$1$ and sensor~$3$, from the example system,~\eqref{ex_sys1}$-$\eqref{ex_sys2}.}
\label{Fl_fig}
\end{figure}

{\bf Dimension of the sub-systems:} The local states at sensor~$l$, i.e., the components of the local state vector,~$\mathbf{x}_k^{(l)}$, are the elements in its associated cut-point set,~$V^{(l)}$. The dimension of the local Kalman filter implemented at sensor $l$ is now $n_l$. The set of $n_l$-dimensional local Kalman filters will give rise, as will be clear later, to an $L$-banded centralized Information matrix with $L=\min(n_1,\ldots,n_N)$. The loss in the optimality as a function of $L$ is given by the divergence~\eqref{divL}. Hence, for a desired level of performance, i.e., for a fixed~$L$, we may need to extend (include additional states in a cut-point set) the cut-point sets,~$V^{(l)}$, to~$V_L^{(l)}$, such that\begin{equation}n_l=\left|V^{(l)}_L\right| \geq L,\qquad\forall~l,\end{equation} where $|\cdot|$ when applied to a set denotes its cardinality. This procedure of choosing an $L$ based on a certain desired performance gives a lower bound on the dimensions of the sub-systems.

{\bf Coupled States as Inputs:} The directed edges coming into a cut-point set are the inputs required by the reduced model. In the context of our running illustration~\eqref{ex_sys1}$-$\eqref{ex_sys2}, we see that the local state vector for sensor~$1$ is,~$\mathbf{x}_k^{(1)}=[x_{1,k},x_{2,k},x_{3,k}]^T$, and the inputs to the local model consist of a subset of the state set,~$X$, (at sensor~$s_1$,~$x_{4,k}$ is the input coming from sensor~$s_2$) and a subset of the noise input set,~$U$, ($u_{2,k}$ at sensor~$s_1$).

{\bf Local models:} For the local model at sensor~$l$, we collect the states required as input in a local internal input vector,~$\mathbf{d}_k^{(l)}$ (we use the word {\it internal} to distinguish from the externally applied inputs), and the noise sources required as input in a local noise input vector,~$\mathbf{u}_k^{(l)}$. We collect the elements from $\mathbf{F}$ corresponding to the local state vector,~$\mathbf{x}_k^{(l)}$, in a local model matrix,~$\mathbf{F}^{(l)}$. Similarly, we collect the elements from $\mathbf{F}$ corresponding to the local internal input vector,~$\mathbf{d}_k^{(l)}$, in a local internal input matrix,~$\mathbf{D}^{(l)}$, and the elements from $\mathbf{G}$ corresponding to the local noise input vector,~$\mathbf{u}_k^{(l)}$, in a local state noise matrix,~$\mathbf{G}^{(l)}$. Figure~\ref{Fl_fig} shows this partitioning for sensors~$s_1$ and~$s_3$. We have the following local models for~\eqref{ex_sys1}. \begin{eqnarray}\label{rm1}\mathbf{x}_{k+1}^{(1)}&=&\left[
\begin{array}{ccc}
f_{11} & f_{12} & 0     \\
f_{21} &     f_{22}  & 0     \\
f_{31} &     0  & f_{33}
\end{array}%
\right]\mathbf{x}_k^{(1)}+\left[
\begin{array}{c}
0\\
f_{24}\\
0
\end{array}%
\right]x_{4,k}+\left[
\begin{array}{c}
0\\
0\\
g_{32}\end{array}%
\right]u_{2,k},\nonumber \\&=&\mathbf{F}^{(1)}\mathbf{x}_k^{(1)}+\mathbf{D}^{(1)}\mathbf{d}_k^{(1)}+\mathbf{G}^{(1)}\mathbf{u}_k^{(1)}\\\nonumber&&\ \\ \label{rm2}\mathbf{x}_{k+1}^{(2)}&=&\left[
\begin{array}{ccc}
f_{22} & 0 & f_{42}     \\
0 & f_{33}  & 0     \\
0 & f_{43}  & 0
\end{array}%
\right]\mathbf{x}_k^{(2)}+\left[
\begin{array}{cc}
f_{21} & 0\\
f_{31} & 0\\
0 & f_{45}
\end{array}%
\right]\left[
\begin{array}{cc}
x_{1,k}\\
x_{5,k}
\end{array}%
\right]+\left[
\begin{array}{c}
0\\
g_{32}\\
0\end{array}%
\right]u_{2,k},\nonumber \\&=&\mathbf{F}^{(2)}\mathbf{x}_k^{(2)}+\mathbf{D}^{(2)}\mathbf{d}_k^{(2)}+\mathbf{G}^{(2)}\mathbf{u}_k^{(2)}\\\nonumber&&\ \\ \label{rm3}\mathbf{x}_{k+1}^{(3)}&=&\left[
\begin{array}{cc}
0 & f_{45}     \\
f_{54}  & f_{55}
\end{array}%
\right]\mathbf{x}_k^{(3)}+\left[
\begin{array}{c}
f_{43}\\
0
\end{array}%
\right]x_{3,k}+\left[\begin{array}{c}0\\g_{51}\end{array}\right]u_{1,k},\nonumber \\&=&\mathbf{F}^{(3)}\mathbf{x}_k^{(3)}+\mathbf{D}^{(3)}\mathbf{d}_k^{(3)}+\mathbf{G}^{(3)}\mathbf{u}_k^{(3)}\end{eqnarray}

We may also capture the above extraction of the local states by the cut-point sets, with the following procedure. Let the total number of states in the cut-point set at sensor $l$,~$V^{(l)}$, be~$n_l$. Let~$\mathbf{T}_l$ be an~$n_l\times n$ selection matrix, such that it selects~$n_l$ states in the cut-point set,~$V^{(l)}$, from the entire state vector,~$\mathbf{x}_k$, according to the following relation,\begin{equation}\label{Tl}\mathbf{x}_k^{(l)}=\mathbf{T}_l\mathbf{x}_k.\end{equation} For sensor~$1$, the selection matrix,~$\mathbf{T}_1$, is \begin{equation}\mathbf{T}_1 = \left[\begin{array}{ccccc}1&0&0&0&0\\0&1&0&0&0\\0&0&1&0&0\end{array}\right].\end{equation}We establish a {\it reduced} local observation matrix,~$\mathbf{H}^{(l)}$, by retaining the terms corresponding to the local state vector,~$\mathbf{x}_k^{(l)}$, from the local observation matrix,~$\mathbf{H}_l$. We may write \begin{equation}\label{locHredH}\mathbf{H}^{(l)}=\mathbf{H}_l\mathbf{T}_l^\#,\end{equation} where `$\#$' denotes the pseudo-inverse of the matrix. In the context of the running illustration, the reduced local observation matrix~$\mathbf{H}^{(1)}=[1,1,1]$ is obtained from the local observation matrix~$\mathbf{H}_1=[1,1,1,0,0]$. Note that~$\mathbf{H}_l$ picks the states from the global state vector,~$\mathbf{x}_k$, whereas~$\mathbf{H}^{(l)}$ picks the states from the local state vector, $\mathbf{x}_k^{(l)}$. The reduced local observation models are given by \begin{equation}\label{dist_obs} \mathbf{y}_k^{(l)}=\mathbf{H}^{(l)}\mathbf{x}_k^{(l)}+\mathbf{w}_k^{(l)}.\end{equation}

We now make some additional comments. For simplicity of the explanation, we refer to our running example,~\eqref{ex_sys1}$-$\eqref{ex_sys2}. We note that the reduced models at the sensors overlap, as shown by the overlapping cut-point sets in Figure~\ref{cutsets_fig}. Due to this overlap, observations corresponding to the shared states are available at multiple sensors that should be fused. We further note that the reduced model~\eqref{rm1} at sensor~$1$ is coupled to the reduced model~\eqref{rm2} at sensor~$2$ through the state~$x_{4,k}$. The state~$x_{4,k}$ at sensor~$1$ does not appear in the local state vector, i.e.,~$x_{4,k} \notin \mathbf{x}_{k}^{(1)}$. But it is still required as an internal input at sensor~$1$ to preserve the global dynamics. Hence, sensor~$2$ communicates the state~$x_{4,k}$, which appears in its local state vector, i.e.,~$x_{4,k} \in \mathbf{x}_{k}^{(2)}$, to sensor~$1$.  Hence at an arbitrary sensor $l$, we derive the reduced model to be\begin{eqnarray}\label{dist_dyn} \mathbf{x}_{k+1}^{(l)}&=&\mathbf{F}^{(l)}\mathbf{x}_k^{(l)}+\mathbf{D}^{(l)}\mathbf{{d}}_{k}^{(l)}+\mathbf{G}^{(l)}\mathbf{u}_k^{(l)}.\end{eqnarray}
Since the value of the state itself is unknown, sensor~$2$ communicates its estimate, $\widehat{x}^{(2)}_{4,k|k}$, to sensor~$1$. This allows sensor~$1$ to complete its local model and preserve global dynamics, thus, taking into account the coupling between the local reduced-order models. This process is repeated at all sensors. Hence, the local internal input vector,~$\mathbf{d}_k^{(l)}$, is replaced by its estimate,~$\mathbf{\widehat{d}}_{k|k}^{(l)}$. It is worth mentioning here that if the dynamics were time-dependent, i.e., the matrices, $\mathbf{F,G,H}$, change with time, $k$, then the above decomposition procedure will have to be repeated~at~each~$k$. This may result into a different communication topology over which the sensors communicate at each $k$.\begin{figure}
\centering
\includegraphics[width=3.5in]{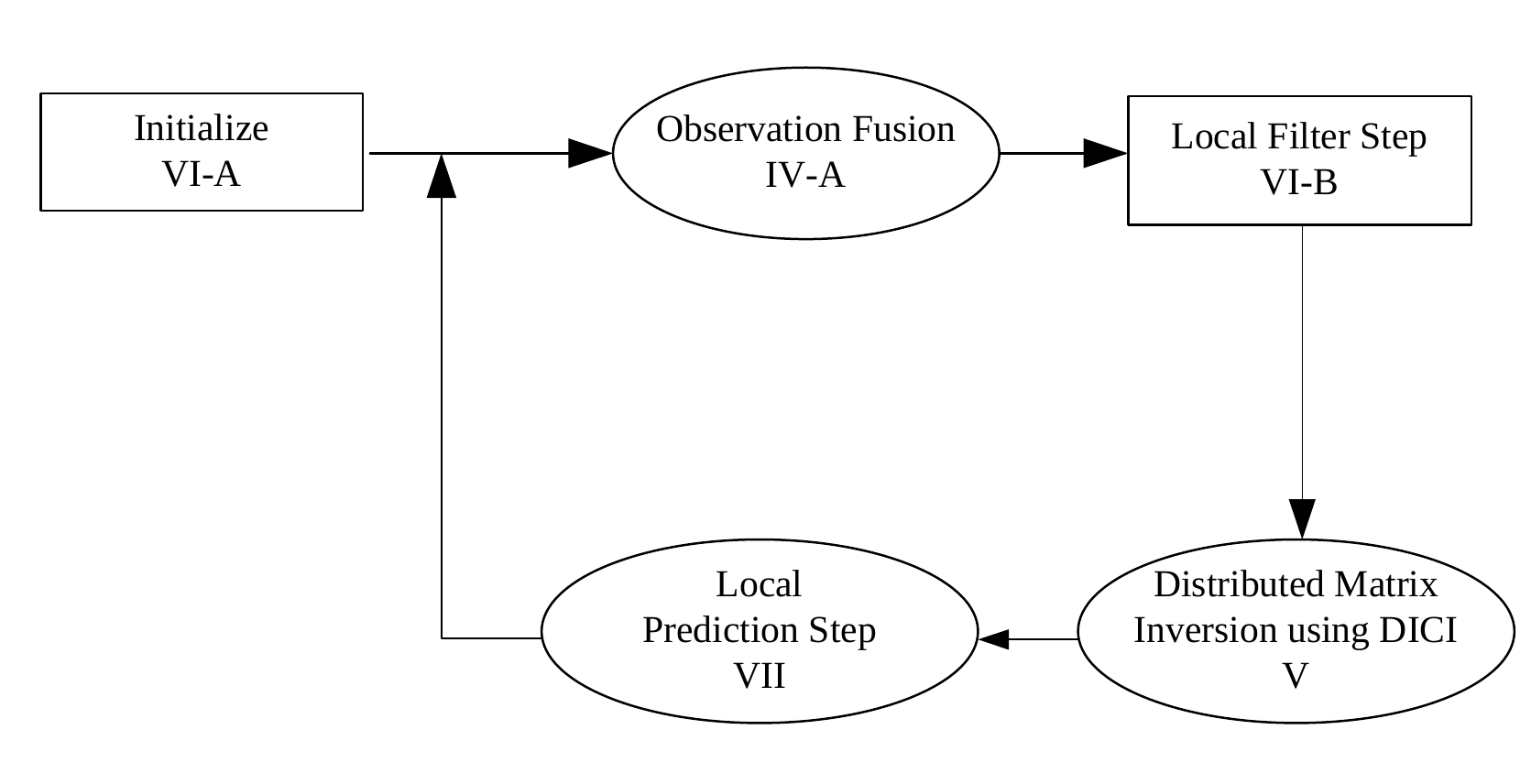}
\caption{Block Diagram for the LIFs: Steps involved in the LIF implementation. The ovals represent the steps that require local communication.}
\label{block_diag}
\end{figure}

\subsection{Local Information Filters}\label{LIF:sub_mod}
To distribute the estimation of the global state vector,~$\mathbf{x}_k$, we implement local Information filters (LIFs) at each sensor~$l$, which are based on the sensor-based reduced models~\eqref{dist_dyn} and~\eqref{dist_obs}. Each LIF computes local objects (matrices and vectors of dimension~$n_l$,) which are then fused (if required) by exchanging information among the neighbors. Some of the update procedures are iterative. Although, no centralized knowledge of the estimation of the global state exists, the union of the local state vector represents, in a distributed way, the knowledge that exists in a centralized fashion in the CLBIF. In most applications nowhere in the network is there the need for this centralized knowledge.

The LIFs consist of initial conditions, a local filter step (including observation fusion and distributed matrix inversion) and a local prediction step (including estimate fusion), see Figure~\ref{block_diag}. These steps are presented in the next four sections. To proceed with the next sections, we provide notation in the following Subsection.

\subsubsection{Notation}\label{LIF:not}
\begin{figure}
\centering
\includegraphics[height=1.25in]{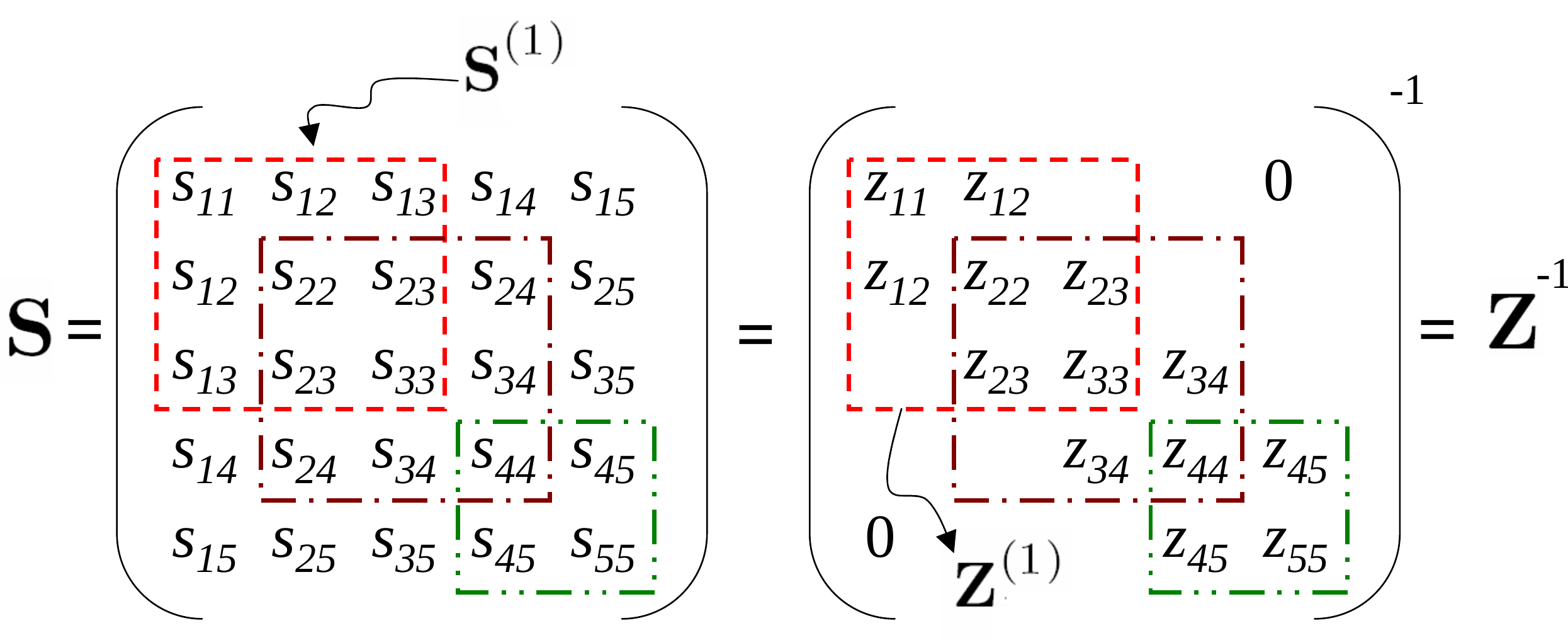}
\caption{Relationship between the global error covariance matrices,~$\mathbf{S}$ and their inverses, the global information matrices,~$\mathbf{Z}$, with~$L=1-$banded approximation on~$\mathbf{Z}$. The figure also shows how the local matrices,~$\mathbf{S}^{(l)}$ and~$\mathbf{Z}^{(l)}$, constitute their global counterparts. Since this relation holds for both the estimation and prediction matrices, the figure removes the subscripts.}
\label{Z2S_fig}
\end{figure}

We use the notation that the superscript~$(l)$ refers to a local reduced-order variable ($n_l\times 1$ vector or~$n_l\times n_l$ matrix) at sensor~$l$ to define the reduced observation vector,~$\mathbf{i}_{k}^{(l)}$, and the reduced~observation~matrix,~$\mathbf{\mathcal{I}}^{(l)}$,~as \begin{eqnarray}\label{red_obs_vec}
\mathbf{i}_{k}^{(l)}&=&(\mathbf{H}^{(l)})^{T}\mathbf{R}_{l}^{-1}\mathbf{y}_{k}^{(l)},\\ \label{red_obs_mat}
\mathbf{\mathcal{I}}^{(l)}&=&(\mathbf{H}^{(l)})^{T}\mathbf{R}_{l}^{-1}\mathbf{H}^{(l)}.\end{eqnarray} The local error covariance matrices,~$\mathbf{S}^{(l)}_{k|k}$ and $\mathbf{S}^{(l)}_{k|k-1}$, are the overlapping diagonal submatrices of the global error covariance matrices,~$\mathbf{S}_{k|k}$ and~$\mathbf{S}_{k|k-1}$. Let $\mathbf{Z}^{(l)}_{k|k}$ and~$\mathbf{Z}^{(l)}_{k|k-1}$ be the local information matrices. These local information matrices are overlapping diagonal submatrices of the global~$L-$banded information matrices,~$\mathbf{Z}_{k|k}$ and~$\mathbf{Z}_{k|k-1}$. These local matrices overlap because the reduced sensor-based models \eqref{dist_dyn} have overlapping state vectors,~$\mathbf{x}_k^{(l)}$. Figure~\ref{Z2S_fig} captures the relationship between the local error covariance matrices and the local information matrices given by~\eqref{Z2S} and \eqref{Z2Pi}.

\section{Overlapping Reduced Models}\label{LIF:orm}
After the model distribution step introduced in section~\ref{mod_dist}, the reduced models among the sensors may share states, as shown by the overlapped cut-point sets in Figure~\ref{cutsets_fig}. Since the sensors sharing the states have independent observations of the shared states, observations corresponding to the shared states should be fused. We present observation fusion in subsection~\ref{obs_fus} with the help of bipartite fusion graphs,~\cite{usman_sspw:07}.

\subsection{Observation Fusion}\label{obs_fus}
Equations~\eqref{ikf} and~\eqref{Ikf} show that the observation fusion is equivalent to adding the corresponding~$n-$dimensional local observation variables,~\eqref{ilk}$-$\eqref{Ilk}. In CLBIF, we implement this fusion directly because each local observation variable in~\eqref{ilk}$-$\eqref{Ilk} corresponds to the full~$n-$dimensional state vector,~$\mathbf{x}_k$. Since the~$n_l-$dimensional \emph{reduced} observation variables,~\eqref{red_obs_vec}$-$\eqref{red_obs_mat}, correspond to different local state vectors,~$\mathbf{x}_k^{(l)}$, they cannot be added directly.

\begin{figure}
\centering
\includegraphics[width=2in]{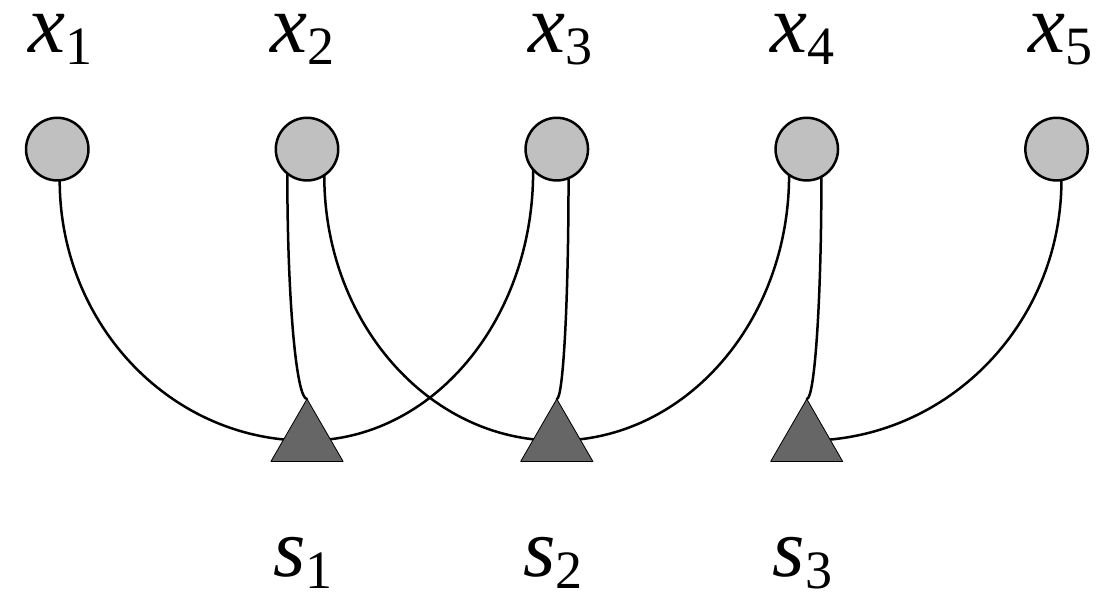}
\caption{Bipartite Fusion graph,~$\mathcal{B}$, is shown for the example system,~\eqref{ex_sys1}$-$\eqref{ex_sys2}.}
\label{bipart}
\end{figure}

To achieve observation fusion, we introduce the following undirected bipartite fusion graph\footnote{A bipartite graph is a graph whose vertices can be divided into two disjoint sets $X$ and $S_N$, such that, every edge connects a vertex in $X$ to a vertex in $S_N$, and, there is no edge between any two vertices of the same set, \cite{bela_book}.},~$\mathcal{B}$. Let~$S_N=\{s_1,\ldots,s_N\}$ be the set of sensors and~$X$ be the set of states. The vertex set of the bipartite fusion graph,~$\mathcal{B}$, is~$S_N \bigcup X$. We now define the edge set,~$E_\mathcal{B}$, of the fusion graph,~$\mathcal{B}$. The sensor~$s_i$ is connected to the state variable~$x_j$, if~$s_i$ observes the state variable~$x_j$. In other words, we have an edge between sensor~$s_i$ and state variable~$x_j$, if the local observation matrix,~$\mathbf{H}_i$, at sensor~$s_i$, contains a non-zero entry in its~$j$th column. Figure~\ref{bipart} shows the bipartite graph for the example system in~\eqref{ex_sys1}$-$\eqref{ex_sys2}.

We now further provide some notation on the communication topology. Let~$\mathcal{G}$ denote the sensor communication graph that determines how the sensors communicate among themselves. Let~$\mathcal{K}(l)$ be the subgraph of~$\mathcal{G}$, that contains the sensor~$l$ and the sensors directly connected to the sensor $l$. For the~$j$th state,~$x_j$, let~$\mathcal{G}_j$ be the induced subgraph of the sensor communication graph,~$\mathcal{G}$, such that it contains the sensors having the state~$x_j$ in their reduced models.
%Note here the difference between the state-dependency graph,~$\mathcal{J}$, that captures the dynamic behavior of the system given by model matrices,~$\mathbf{F\mbox{ and }G}$, the sensor communication graph,~$\mathcal{G}$, that captures how the sensors communicate with each other, and the bipartite fusion graph,~$\mathcal{B}$, defined above.

The set of vertices in~$\mathcal{G}_j$ come directly form the bipartite fusion graph,~$\mathcal{B}$. For example, from Figure~\ref{bipart}, we see that $\mathcal{G}_1$ contains~$s_1$ as a single vertex,~$\mathcal{G}_2$ contains~$s_1, s_2$ as vertices, and so on. States having more than one sensor connected to them in the bipartite fusion graph,~$\mathcal{B}$, are the states for which fusion is required, since we have multiple observations for that state. Furthermore, Figure~\ref{bipart} also gives the vertices in the associated subgraphs,~$\mathcal{G}_j$, over which the fusion is to be carried out.

With the help of the above discussion, we establish the fusion of the reduced observation variables,~\eqref{red_obs_vec}$-$\eqref{red_obs_mat}. The reduced model at each sensor involves~$n_l$ state variables, and each element in the~$n_l\times 1$ reduced observation vector,~$\mathbf{i}_k^{(l)}$, corresponds to one of these states, i.e., each entry in~$\mathbf{i}_k^{(l)}$ has some information about its corresponding state variable. Let the entries of the~$n_l\times 1$ reduced observation vector,~$\mathbf{i}_k^{(l)}$, at sensor~$l$, be subscripted by the~$n_l$ state variables modeled at sensor~$l$. In the context of the example given by system~\eqref{ex_sys1}$-$\eqref{ex_sys2}, we have \begin{align}\mathbf{i}_k^{(1)}&=\left[\begin{array}{c}i^{(1)}_{k,x_1}\\i^{(1)}_{k,x_2}\\i^{(1)}_{k,x_3}\end{array}\right],&
\mathbf{i}_k^{(2)}&=\left[\begin{array}{c}i^{(2)}_{k,x_2}\\i^{(2)}_{k,x_3}\\i^{(2)}_{k,x_4}\end{array}\right],&
\mathbf{i}_k^{(3)}&=\left[\begin{array}{c}i^{(3)}_{k,x_4}\\i^{(3)}_{k,x_5}\end{array}\right].\end{align} For each state~$x_j$, the observation fusion is carried out on the sensors attached to this state in the bipartite fusion graph,~$\mathcal{B}$. The fused observation vectors denoted by~$\mathbf{i}_{f,k}^{(l)}$ are given by \begin{align}\label{fusi}\mathbf{i}_{f,k}^{(1)}&=\left[\begin{array}{c}i^{(1)}_{k,x_1}\\i^{(1)}_{k,x_2}+i^{(2)}_{k,x_2}\\i^{(1)}_{k,x_3}+i^{(2)}_{k,x_3}\end{array}\right],&
\mathbf{i}_{f,k}^{(2)}&=\left[\begin{array}{c}i^{(2)}_{k,x_2}+i^{(1)}_{k,x_2}\\i^{(2)}_{k,x_3}+i^{(1)}_{k,x_3}\\i^{(2)}_{k,x_4}+i^{(3)}_{k,x_4}\end{array}\right],&
\mathbf{i}_{f,k}^{(3)}&=\left[\begin{array}{c}i^{(3)}_{k,x_4}+i^{(2)}_{k,x_4}\\i^{(3)}_{k,x_5}\end{array}\right].\end{align} Generalizing to the arbitrary sensor~$l$, we may write the entry,~$i^{(l)}_{f,k,{x_j}}$, corresponding to~$x_j$ in the fused observation vector,~$\mathbf{i}_{f,k}^{(l)}$ as \begin{equation}\label{o_f_waa}i_{f,k,{x_j}}^{(l)}=\sum_{s\in \mathcal{G}_j} i_{k,x_j}^{(s)},\end{equation} where~$i_{k,x_j}^{(s)}$ is the entry corresponding to $x_j$ in the reduced observation vector at sensor~$s$,~$\mathbf{i}_k^{(s)}$.

Since the communication network on~$\mathcal{G}_j$ may not be all-to-all, an iterative weighted averaging algorithm~\cite{boyd:04}, is used to compute the fusion in \eqref{o_f_waa} over arbitrarily connected communication networks with only local communication (the communication may be multi-hop). A similar procedure on the pairs of state variables and their associated subgraphs can be implemented to fuse the reduced observation matrices,~$\mathbf{\mathcal{I}}^{(l)}$. Since, we assume the observation model to be stationary ($\mathbf{H}$ and $\mathbf{R}$ are time-independent), the fusion on the reduced observation matrix,~$\mathbf{\mathcal{I}}^{(l)}$, is to be carried out only once and can be an offline procedure. If that is not the case and $\mathbf{H}$ and $\mathbf{R}$ are time dependent fusion on $\mathbf{\mathcal{I}}$ has to be repeated at each~time,~$k$.

Here, we assume that the communication is fast enough so that the consensus algorithm can converge, see \cite{zamp:07} for a discussion on distributed Kalman filtering based on consensus strategies. The convergence of the consensus algorithm is shown to be geometric and the the convergence rate can be increased by optimizing the weight matrix for the consensus iterations using semidefinite programming (SDP)~\cite{xiao:04}. The communication topology of the sensor network can also be improved to increase the convergence speed of the consensus algorithms~\cite{saeed:05,kar:06,kar1:07}.

Here, we comment on estimate fusion. Since we fused the observations concerning the shared states among the sensors, one may ask if it is required to carry out fusion of the estimates of the shared states. It turns out that consensus on the observations leads to consensus on the estimates. This will become clear with the introduction of the local filter and the local prediction step of the LIFs, therefore, we defer the discussion on estimate fusion to section~\ref{est_fus}.

\section{Distributed Matrix Inversion with Local Communication}\label{dist_jac}
In this section, we discuss the fusion of the error covariances. Consider the example model~\eqref{ex_sys1}$-$\eqref{ex_sys2}, when we employ LIFs on the distributed models~\eqref{rm1}$-$\eqref{rm3}. The local estimation information matrices,~$\mathbf{Z}_{k|k}^{(1)}$,~$\mathbf{Z}_{k|k}^{(2)}$, and~$\mathbf{Z}_{k|k}^{(3)}$, correspond to the overlapping diagonal submatrices of the global~$5\times 5$ estimation information matrix,~$\mathbf{Z}_{k|k}$, see Figure~\ref{Z2S_fig}, with~$L=1-$banded assumption on~$\mathbf{Z}_{k|k}$. It will be shown (section~\ref{PD_1}) that the local prediction information matrix,~$\mathbf{Z}_{k+1|k}^{(l)}$, is a function of the local error covariance matrices,~$\mathbf{S}_{k|k}^{(l)}$, and hence we need to compute~$\mathbf{S}_{k|k}^{(l)}$ from the local filter information matrices, $\mathbf{Z}_{k|k}^{(l)}$, which we get from the local filter step (section~\ref{LIF:ic_lfs}). As can be seen from Figure~\ref{Z2S_fig} and~\eqref{Z2S}, for these local submatrices, \begin{equation}\label{Zn2S}{\mathbf{S}^{(l)}}\neq\left(\mathbf{Z}^{(l)}\right)^{-1}.\end{equation}

Collecting all the local information matrices,~$\mathbf{Z}_{k|k}^{(l)}$, at each sensor and then carrying out an~$n\times n$ matrix inversion is \emph{not} a practical solution for large-scale systems (where~$n$ may be large), because of the large communication overhead and~$O(n^3)$ computational cost. Using the~$L-$banded structure on the global estimation information matrix,~$\mathbf{Z}_{k|k}$, we present below a distributed iterate collapse inversion overrelaxation (DICI-OR, pronounced {\it die-see}--O--R) algorithm\footnote{It is worth mentioning here that the DICI algorithm (for solving $\mathbf{ZS=I}_{n\times n},$ with SPD $L-$banded matrix $\mathbf{Z}\in\mathbb{R}^{n\times n}$ and the $n\times n$ identity matrix $\mathbf{I}_{n\times n}$) is neither a direct extension nor a generalization of (block) Jacobi or Gauss-Seidel type iterative algorithms (that solve a vector version, $\mathbf{Zs=b}$ with $\mathbf{s,b}\in\mathbb{R}^n$, of $\mathbf{ZS=I}_{n\times n}$, see~\cite{tsit_book,garloff:90,siljak:94,zev:95,baraniuk:06}.) Using the Jacobi or Gauss-Seidel type iterative schemes for solving $\mathbf{ZS=I}$ is equivalent to solving $n$ linear systems of equations, $\mathbf{Zs=b}$; hence, the complexity scales linearly with $n$. Instead the DICI algorithm employs a non-linear collapse operator that exploits the structure of the inverse, $\mathbf{S}$, of a symmetric positive definite $L-$banded matrix, $\mathbf{Z}$, which makes its complexity independent of $n$.}%\footnote{Forward-Backward Recursion Algorithm: For an~$L-$banded matrix,~$\mathbf{Z}_{k|k}$, a forward-backward recursion algorithm to invert $\mathbf{Z}_{k|k}$ using only the submatrices in its~$L-$band is provided in~\cite{kavcic:00}. This algorithm computes the local error covariance matrices, $\mathbf{S}^{(l)}_{k|k}$, from the local information matrices,~$\mathbf{Z}^{(l)}_{k|k}$. The problem in implementing this in a multisensor environment is that it is sequential: involving a forward recursion to start from the first sensor and reach the last sensor, and a backward recursion that proceeds in the opposite direction. Since, the iterations involve serial communication of the local matrices among all the sensors, the associated latency is impractical, besides requiring an inordinate amount of communication.}
. For notational convenience, we disregard the time indices in the following development.

We present a generalization of the centralized Jacobi overrelaxation (JOR) algorithm to solve matrix inversion in section~\ref{ja} and show that the computations required in its distributed implementation scale linearly with the dimension, $n$, of the system. We then present the DICI-OR algorithm and show that it is independent of the dimension, $n$, of the system.
%We, further, illustrate the DICI-OR algorithm with the help of an example in section~\ref{gdj}.

\subsection{Centralized Jacobi Overrelaxation (JOR) Algorithm}\label{ja}
The centralized JOR algorithm for vectors~\cite{tsit_book} solves a linear system of~$n$ equations iteratively, by successive substitution. It can be easily extended to get the centralized JOR algorithm for matrices that solves \begin{equation}\label{ZST}\mathbf{ZS=T},\end{equation} for the unknown matrix, $\mathbf{S}$, where the matrices~$\mathbf{Z}\mbox{ and }\mathbf{T}$ are known. Let~$\mathbf{M}= \mbox{diag} (\mathbf{Z})$, for some~$\gamma>0$\begin{equation}\label{jac_iter_mat}
\mathbf{S}_{t+1}=\left(\left(1-\gamma\right)\mathbf{I}_{n\times n}+\gamma\mathbf{M}^{-1}\left(\mathbf{M -Z}\right)\right)\mathbf{S}_{t}+\gamma\mathbf{M^{-1}T},
\end{equation} converges to~$\mathbf{S}$, and is the centralized JOR algorithm for matrices, solving~$n$ coupled linear systems of equations~\eqref{ZST}, where $\gamma$ is sometimes called a relaxation parameter~\cite{tsit_book}. Putting~$\mathbf{T=I}_{n\times n}$, we can solve for $\mathbf{ZS=I}_{n\times n}\Rightarrow\mathbf{S=Z}^{-1}$, and, if~$\mathbf{Z}$ is known, the following iterations converge to~$\mathbf{Z}^{-1}$, \begin{equation}\label{jac_or}
\mathbf{S}_{t+1}=\mathbf{P}_\gamma\mathbf{S}_{t}+\gamma\mathbf{M^{-1}},
\end{equation} where the multiplier matrix, $\mathbf{P}_\gamma$, is defined as \begin{equation}\label{P}\mathbf{P}_\gamma=\left(1-\gamma\right)\mathbf{I}_{n\times n}+\gamma \mathbf{M}^{-1}\left(\mathbf{M-Z}\right).\end{equation} The Jacobi algorithm can now be considered as a special case of the JOR algorithm with $\gamma=1$.

\subsubsection{Convergence}\label{convjor}
Let~$\mathbf{S}$ be the stationary point of the iterations in~\eqref{jac_or}, and let~$\mathbf{\widetilde{S}}_{t+1}$ denote the iterations of~\eqref{jac_or}. It can be shown that the error process,~$\mathbf{\widetilde{E}}_{t+1}$, for the JOR algorithm is
\begin{eqnarray}\label{E_DJ}\mathbf{\widetilde{E}}_{t+1}&=&\mathbf{P_\gamma}\mathbf{\widetilde{E}}_{t},\end{eqnarray}
which decays to zero if~$||\mathbf{P}_\gamma||_2<1$, where $||\cdot||_2$ denotes the spectral norm of a matrix. The JOR algorithm \eqref{jac_or} converges for all symmetric positive definite matrices, $\mathbf{Z}$, for sufficiently small $\gamma>0$, see~\cite{tsit_book}, an alternate convergence proof is provided in \cite{siljak:00} via convex $M$-matrices, whereas, convergence for parallel asynchronous team algorithms is provided in \cite{bhaya:96}. Since, the information matrix, $\mathbf{Z}$, is the inverse of an error covariance matrix; the information matrix, $\mathbf{Z}$, is symmetric positive definite by definition and the JOR algorithm always converges. Plugging $\gamma=1$ in \eqref{jac_or} gives us the centralized Jacobi algorithm for matrices, which converges for all diagonally dominant matrices, $\mathbf{Z}$, see \cite{tsit_book}. We can further write~the~error~process~as, \begin{eqnarray}\mathbf{\widetilde{E}}_{t+1}&=& \mathbf{P}_\gamma^{t+1}\left(\mathbf{S}_0-\mathbf{S}\right),\end{eqnarray} where the matrix $\mathbf{S}_0$ is the initial condition. The spectral norm of the error process can be bounded by \begin{eqnarray}\label{err_bnd}||\mathbf{\widetilde{E}}_{t+1}||_2 &\leq&|\lambda_{\max}(\mathbf{P_\gamma})|^{t+1}~|| \left(\mathbf{S}_0-\mathbf{S}\right)||_2,\end{eqnarray} where $\lambda_{\max}$ is the maximum eigenvalue (in magnitude) of the multiplier matrix, $\mathbf{P}_\gamma$.
The JOR algorithm is centralized as it requires the complete~$n\times n$ matrices involved. This requires global communication and an~$n$th order computation at each iteration of the algorithm. We present below its distributed implementation.

\subsubsection{Distributed JOR algorithm}\label{djor}
We are interested in the local error covariances that lie on the $L-$band of the matrix $\mathbf{S=Z}^{-1}$. Distributing the JOR (in addition to \cite{tsit_book}, distributed Jacobi and Gauss Seidel type iterative algorithms can also be found in \cite{garloff:90,zev:95,siljak:00}) algorithm \eqref{jac_or} directly to compute the $L-$band of $\mathbf{S}$ gives us the following equations for the $ij$-th element, $s_{ij}$, in $\mathbf{S}_{t+1}$,\begin{eqnarray}\label{dj1}s_{ij,t+1} &=& \mathbf{p}_i\mathbf{s}_t^j,~~~\qquad\qquad\qquad i\neq j,\\\label{dj2}s_{ij,t+1} &=& \mathbf{p}_i\mathbf{s}_t^j+m_{ii}^{-1},\qquad\qquad i=j,\end{eqnarray} where: the row vector, $\mathbf{p}_i$, is the $i$th row of the multiplier matrix, $\mathbf{P}_\gamma$; the column vector, $\mathbf{s}_t^j$, is the $j$th column of the matrix, $\mathbf{S}_t$; and the scalar element, $m_{ii}$, is the $i$th diagonal element of the diagonal matrix, $\mathbf{M}$. Since the matrix $\mathbf{Z}$ is $L-$banded, the multiplier matrix, $\mathbf{P}_\gamma$, in \eqref{P} is also $L-$banded. The $i$th row of the multiplier matrix, $\mathbf{P}_\gamma$, contains non-zeros at most at $2L+1$ locations in the index set, $\kappa=\{i-L,\ldots,i,\ldots,i+L\}$. These non-zero elements pick the corresponding elements with indices in the index set, $\kappa$, in the $j$th column, $\mathbf{s}^j_t$, of $\mathbf{S}_t$. Due to the $L-$bandedness of the multiplier matrix, $\mathbf{P}_\gamma$, the JOR algorithm can be easily distributed with appropriate communication with the neighboring sensors.

A major drawback of the distributed JOR algorithm is that at each sensor the computation requirements scale linearly with the dimension, $n$, of the system. This can be seen by writing out the iteration in \eqref{dj1}, e.g., for an $L-$banded element such that $|i-j|=L$. In the context of Figure~\ref{Z2S_fig}, we can write $s_{45,t+1}$ from~\eqref{dj1} as \begin{eqnarray}\label{part3}s_{45,t+1} &=& p_{43}s_{35,t} + p_{44}s_{45,t} + p_{45}s_{55,t}\end{eqnarray} The element~$s_{35,t}$ does not lie in the~$L-$band of~$\mathbf{S}$, and, hence, does not belong to any local error covariance matrix,~$\mathbf{S}^{(l)}$. Iterating on it using~\eqref{dj1} gives \begin{equation}\label{s35}s_{35,t+1} = p_{32}s_{25,t} + p_{33}s_{35,t} + p_{34}s_{45,t}.\end{equation} The computation in~\eqref{s35}, involves~$s_{25,t}$, iterating on which, in turn, requires another off~$L-$band element,~$s_{15,t}$, and so on. Hence, a single iteration of the algorithm, although distributed and requiring only local communication, sweeps the entire rows in~$\mathbf{S}$ and the computation requirements scale linearly with $n$. We now present a solution to this problem.

\subsection{Distributed Iterate Collapse Inversion Overrelaxation (DICI-OR) Algorithm}\label{gdj}
In this section, we present the distribute iterate collapse inversion overrelaxation (DICI-OR) algorithm. The DICI-OR algorithm is divided into two steps: (i) an iterate step and (ii) a collapse step. The {\bf iterate step} can be written, in general, for the $ij$-th element that lies in the $L-$band ($|i-j|\leq L$) of the matrix $\mathbf{S}_{t+1}$ as \begin{eqnarray}\label{dici1}s_{ij,t+1} &=& \mathbf{p}_i\mathbf{s}_t^j,~~~\qquad\qquad\qquad i\neq j,~|i-j|\leq L,\\\label{dici2}s_{ij,t+1} &=& \mathbf{p}_i\mathbf{s}_t^j+m_{ii}^{-1},\qquad\qquad i=j,~|i-j|\leq L,\end{eqnarray}where the symbols are defined as in section~\ref{djor}.

As we explained before in section~\ref{djor}, the implementation of \eqref{dici1}--\eqref{dici2} requires non $L-$banded elements that, in turn, require more non $L-$banded elements. To address this problem we introduce a {\bf collapse step}. We assume that~$\mathbf{S}_t$ is the inverse of an~$L-$banded matrix and use the results\footnote{If~$\mathbf{S}$ is the inverse of an~$L-$banded matrix, then the submatrices that do not lie in the $L-$band of~$\mathbf{S}$, can be computed from the submatrices that lie in the~$L-$band of~$\mathbf{S}$,~\cite{asif:05}. So, to compute the inverse,~$\mathbf{S=Z}^{-1}$, of an $L-$banded matrix,~$\mathbf{Z}$, we just compute the submatrices that lie in the~$L-$band of its inverse,~$\mathbf{S}$; the remaining submatrices are derived from these using the expressions given in \cite{asif:05}.} in~\cite{asif:05}, to compute a non~$L-$band element~($s_{ij}$~such that $|i-j|>L$) from the $L-$band elements ($s_{ij}$ such that $|i-j|\leq L$). In general, a non $L-$band element in a matrix whose inverse is $L=1-$banded can be written as \begin{equation}\label{sij_non}s_{ij}=s_{i,j-1}s_{i+1,j-1}^{-1}s_{i+1,j},\qquad\qquad |i-j|>L\end{equation} which gives us the collapse step. In the context of Figure~\ref{Z2S_fig}, instead of iterating on~$s_{35}$ as in~\eqref{s35}, we employ the collapse step, \begin{equation}\label{s35_non}s_{35,t}=s_{34,t}s_{44,t}^{-1}s_{45,t},\end{equation} that prevents us from iterating further on the non $L-$banded elements.

The initial conditions of the DICI-OR algorithm are given by \begin{eqnarray}\label{Pl}\mathbf{P}_\gamma^{(l)}&=&(1-\gamma)\mathbf{I}_{n\times n }^{(l)}+\gamma\left(\mathbf{M}^{(l)}\right)^{-1}\left(\mathbf{M}^{(l)}-\mathbf{Z}^{(l)}\right),\\\label{ic}\mathbf{S}^{(l)}_0 &=& \left(\mathbf{Z}^{(l)}\right)^{-1}\end{eqnarray} Note that equations~\eqref{Pl}--\eqref{ic} do not require any communication and can be computed at each sensor directly from the local information matrix,~$\mathbf{Z}^{(l)}$. This is because the matrix~$\mathbf{M}$ is diagonal, and its local submatrix,~$\mathbf{M}^{(l)}$ in \eqref{Pl}, is the exact inverse of the matrix formed by the diagonal elements of~$\mathbf{Z}^{(l)}$.

We refer to equations~\eqref{dici1}$-$\eqref{sij_non}, combined with ~\eqref{Pl}--\eqref{ic}, and appropriate communication from neighboring sensors as the DICI-OR algorithm. The DICI-OR algorithm can be easily extended to $L>1$. The only step to take care of is the collapse step, since~\eqref{sij_non} holds only for~$L=1$. The appropriate formulae to replace~\eqref{sij_non}, when~$L>1$, are provided in~\cite{asif:05}. The computation requirements for the DICI algorithm are independent of $n$ and DICI provides a scalable implementation of the matrix inversion problem. The DICI algorithm (without the overrelaxation parameter, $\gamma$) can be obtained from the DICI-OR algorithm by setting $\gamma=1$.

\subsubsection{Convergence of the DICI-OR algorithm} The iterate and the collapse step of the DICI algorithm over the entire sensor network can be combined in matrix form as follows. \begin{eqnarray}\label{D1}\mbox{Iterate Step:}\qquad\qquad \mathbf{S}_{t+1} &=& \mathbf{P}_\gamma\overline{\mathbf{S}}_t + \mathbf{M}^{-1},~\qquad\qquad|i-j|\leq L\\\label{D2} \mbox{Collapse Step:}\qquad\qquad \overline{\mathbf{S}}_{t+1}&=&\zeta\left(\mathbf{S}_{t+1}\right),\qquad\qquad\qquad|i-j| > L\end{eqnarray} The operator $\zeta(\cdot)$ is the collapse operator; it takes an arbitrary symmetric positive definite matrix and converts it to a symmetric positive definite matrix whose inverse is $L-$banded by using the results in \cite{asif:05}. The DICI algorithm is a composition of the linear iterate operator, $\mathbf{P}_\gamma$ given in \eqref{P}, followed by the collapse operator, $\zeta(\cdot)$ given in \eqref{sij_non} for $L=1$ and in \cite{asif:05} for $L>1$. Combining \eqref{D1} and \eqref{D2} summarizes the DICI algorithm as, \begin{equation}\mathbf{\overline{S}}=\zeta\left(\mathbf{P}_\gamma\left(\mathbf{\overline{S}}+\left(\mathbf{P}_\gamma\mathbf{M}\right)^{-1}\right)\right).\end{equation} We define a composition map, $\mathbf{\Upsilon} : \mathbf{\Xi}\mapsto\mathbf{\Xi}$, where $\mathbf{\Xi}\subset\mathbb{R}^{n\times n}$ is the set of all symmetric positive definite matrices, as $\mathbf{\mathbf{\Upsilon}}\doteq\zeta\circ\mathbf{P}_\gamma$. To prove the convergence of the DICI-OR algorithm, we are required to show that the composition map, $\mathbf{\Upsilon}$, is a contraction map under some norm, that we choose to be the spectral norm $||\cdot||_2$, i.e., for $\alpha\in [0,1)$, \begin{equation}\label{cntr_map}||\mathbf{\Upsilon}(\mathbf{X}_\mathbf{\Xi})-\mathbf{\Upsilon}(\mathbf{Y}_\mathbf{\Xi})||_2 \leq \alpha ||\mathbf{X}_\mathbf{\Xi}-\mathbf{Y}_\mathbf{\Xi}||_2,\qquad\forall\mathbf{X}_\mathbf{\Xi},\mathbf{Y}_\mathbf{\Xi}\in\mathbf{\Xi}\end{equation} The convergence of the iterate step of the DICI-OR algorithm is based on the iterate operator, $\mathbf{P}_\gamma$, which is proved to be a contraction map in \cite{tsit_book}. For the convergence of the collapse operator, $\zeta$, we resort to a numerical procedure and show, in the following, that \eqref{cntr_map} is a contraction by simulating \eqref{cntr_map} $1.17\times 10^6$ times.

\begin{figure}
\centering
\includegraphics[height=2in]{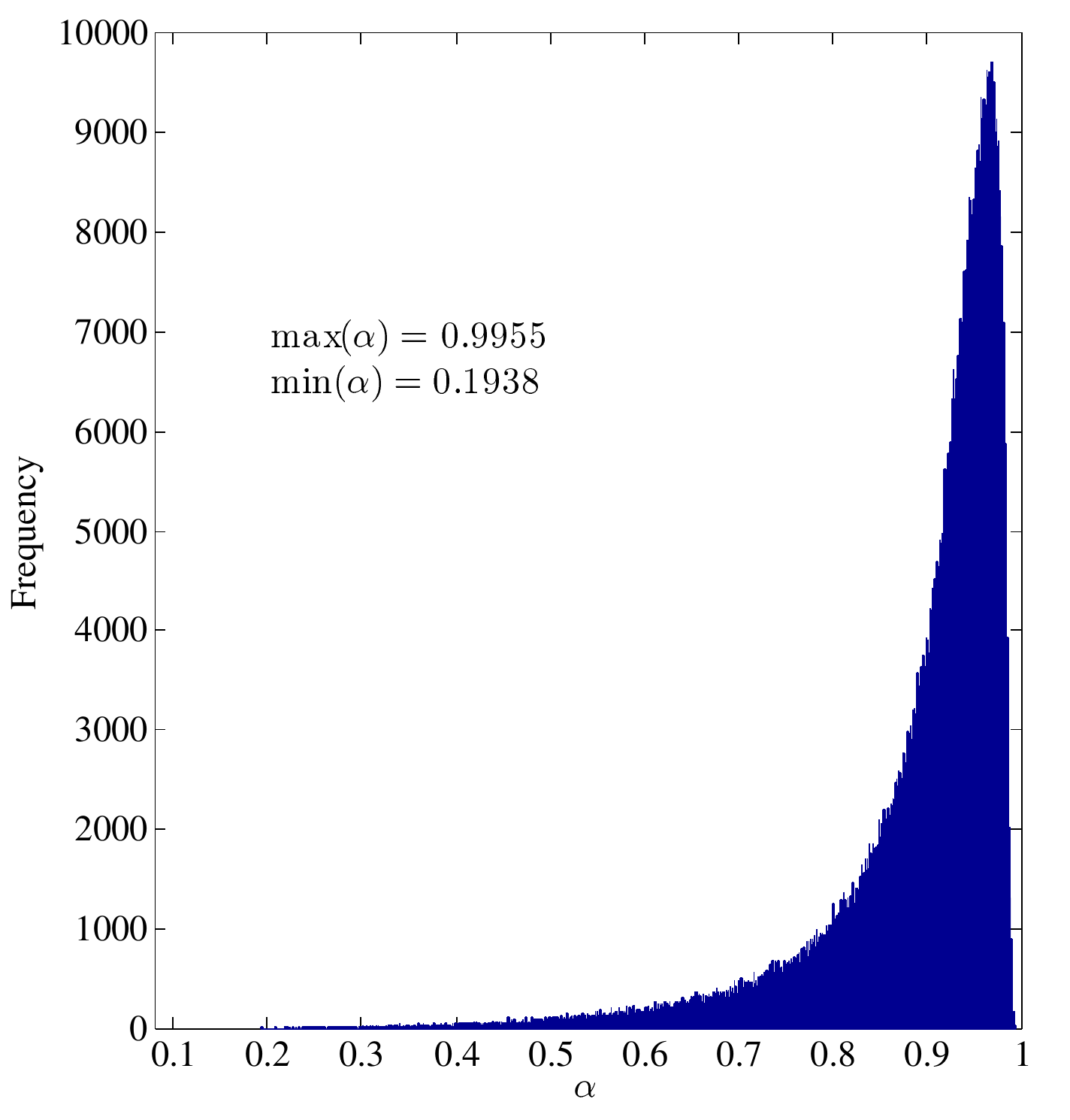}
\caption{Histogram of $\alpha$: Simulations of the quotient of~\eqref{dici_quo} are performed $1.17\times 10^6$ times and the results are provided as a histogram.}
\label{hist_al}
\end{figure}

For the simulations, we generate $n\times n$ matrices, $\mathbf{X}_{\mbox{rand}}$, with i.i.d. normally distributed elements and get, $\mathbf{X}_{\mbox{sym}}=\mathbf{X}_{\mbox{rand}}+\mathbf{X}_{\mbox{rand}}^T$. We eigen-decompose $\mathbf{X}_{\mbox{sym}}=\mathbf{V}\mathbf{\Lambda}\mathbf{V}^T$. We replace $\mathbf{\Lambda}$ with a diagonal matrix, $\mathbf{\Lambda}_{\Xi}$, whose diagonal elements are drawn from a uniform distribution in the interval $(0,10]$. This leads to a random symmetric positive definite matrix, $$\mathbf{X}_{\Xi}=\mathbf{V}\mathbf{\Lambda}_\Xi\mathbf{V}^T.$$ For $n=100$ and $L$ a random integer between $1$ and $n/2=50$, we compute, by Monte Carlo simulations, the quotient of \eqref{cntr_map} \begin{equation}\label{dici_quo}\dfrac{||\mathbf{\Upsilon}(\mathbf{X}_\mathbf{\Xi})-\mathbf{\Upsilon}(\mathbf{Y}_\mathbf{\Xi})||_2} {||\mathbf{X}_\mathbf{\Xi}-\mathbf{Y}_\mathbf{\Xi}||_2}.\end{equation}  The number of trials is $1.17\times 10^6$. The histogram of the values of $\alpha$, in \eqref{dici_quo}, (with $1000$ bins) is plotted in Figure~\ref{hist_al}. The maximum value of $\alpha$ found in these $1.17\times 10^6$ simulations is $0.9955$ and the minimum value is $0.1938$. Since $\alpha\in(0,1)$, i.e., strictly less than 1, we assume that \eqref{cntr_map} is numerically verified.
\subsubsection{Error Bound for the DICI-OR algorithm}
Let the matrix produced by the DICI-OR algorithm at the $t+1$-th iteration be $\mathbf{\widehat{S}}_{t+1}$. The error process in the DICI-OR algorithm is given by \begin{equation}\mathbf{\widehat{E}}_{t+1} = \mathbf{\widehat{S}}_{t+1} - \mathbf{S}.\end{equation}

{\bf Claim:} The spectral norm of the error process, $||\mathbf{\widehat{E}}_{t+1}||_2$, of the DICI-OR algorithm is {\it bounded above} by the spectral norm of the error process, $||\mathbf{\widetilde{E}}_{t+1}||_2$, of the JOR algorithm. Since, the JOR algorithm always converges for symmetric positive definite matrices, $\mathbf{Z}$, we deduce that the DICI-OR algorithm converges.

We verify this claim numerically by Monte Carlo simulations. The number of trials is $4490$, and we compute the error process, $\mathbf{\widehat{E}}_{t+1}(K)$, of the DICI-OR algorithm and the error process, $\mathbf{\widetilde{E}}_{t+1}(K)$, of the JOR algorithm. We choose the relaxation parameter,~$\gamma$, to be~$0.1$. In Figure~\ref{eb1}, Figure~\ref{eb2}, and Figure~\ref{eb3}, we show the following,
\begin{eqnarray*}
\max_K &&\left(||\mathbf{\widetilde{E}}_{t+1}(K)||_2 - ||\mathbf{\widehat{E}}_{t+1}(K)||_2\right),\\
\min_K && \left(||\mathbf{\widetilde{E}}_{t+1}(K)||_2 - ||\mathbf{\widehat{E}}_{t+1}(K)||_2\right),\\
\mbox{mean}_K && \left(||\mathbf{\widetilde{E}}_{t+1}(K)||_2 - ||\mathbf{\widehat{E}}_{t+1}(K)||_2\right),\end{eqnarray*}
respectively, against the number of iterations of the JOR and DICI-OR algorithm. Since all the three figures show that the $\max$, $\min$, and the $\mbox{mean}$ of the difference of the spectral norm of the two error processes, $||\mathbf{\widetilde{E}}_{t+1}(K)||_2 - ||\mathbf{\widehat{E}}_{t+1}(K)||_2$, is always $\geq 0$, we deduce that (using equation~\eqref{err_bnd}) \begin{eqnarray}||\mathbf{\widehat{E}}_{t+1}||_2&\leq&\nonumber||\mathbf{\widetilde{E}}_{t+1}||_2,\\&\leq&|\lambda_{\max}(\mathbf{P_\gamma})|^{t+1}~|| \left(\mathbf{S}_0-\mathbf{S}\right)||_2.\end{eqnarray}

\begin{figure}
\centering
\subfigure[]
{
    \label{eb1.pdf}
    \includegraphics[width=1.8in]{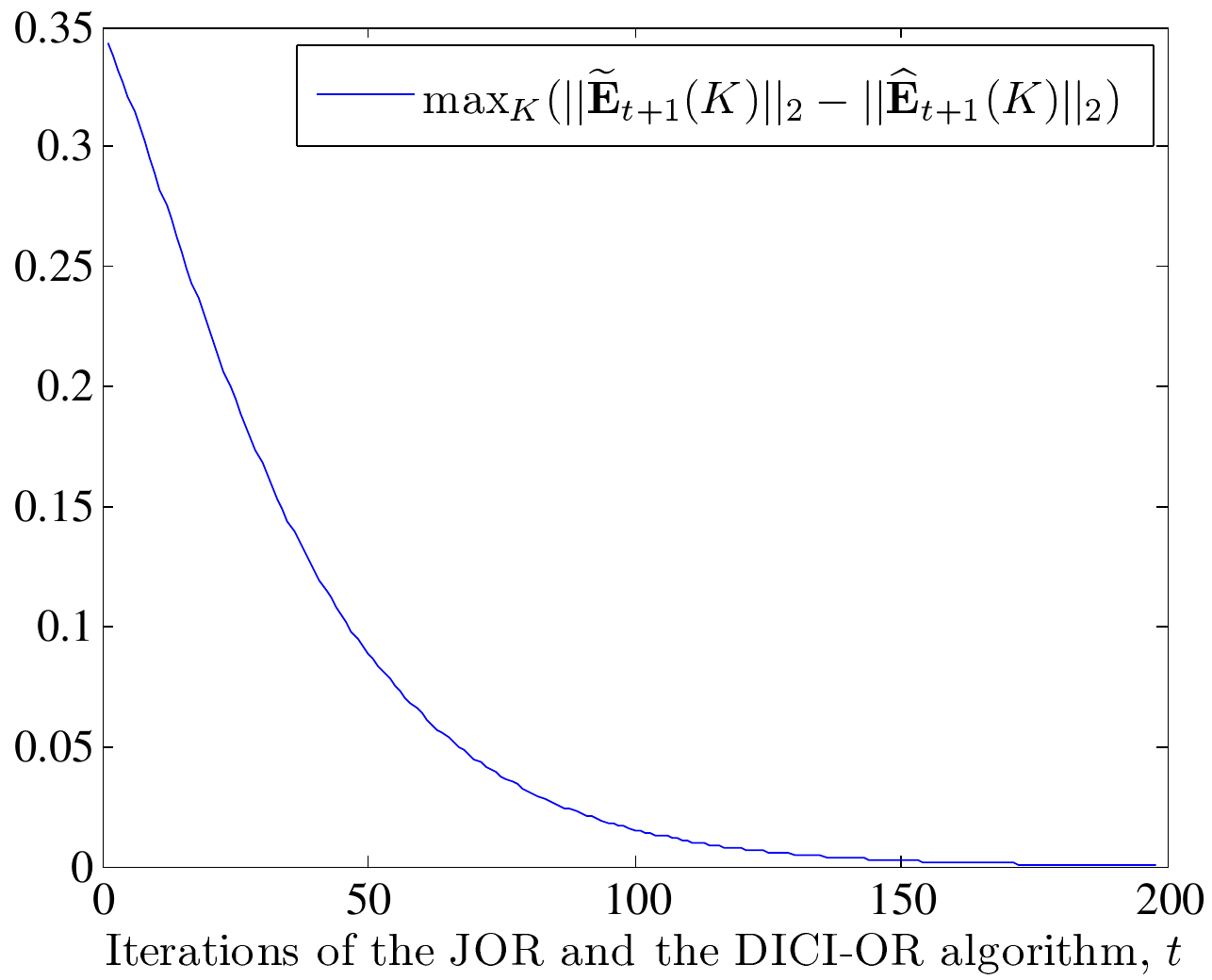}
}
\hspace{.5cm}
\subfigure[]
{
    \label{eb2.pdf}
    \includegraphics[width=1.8in]{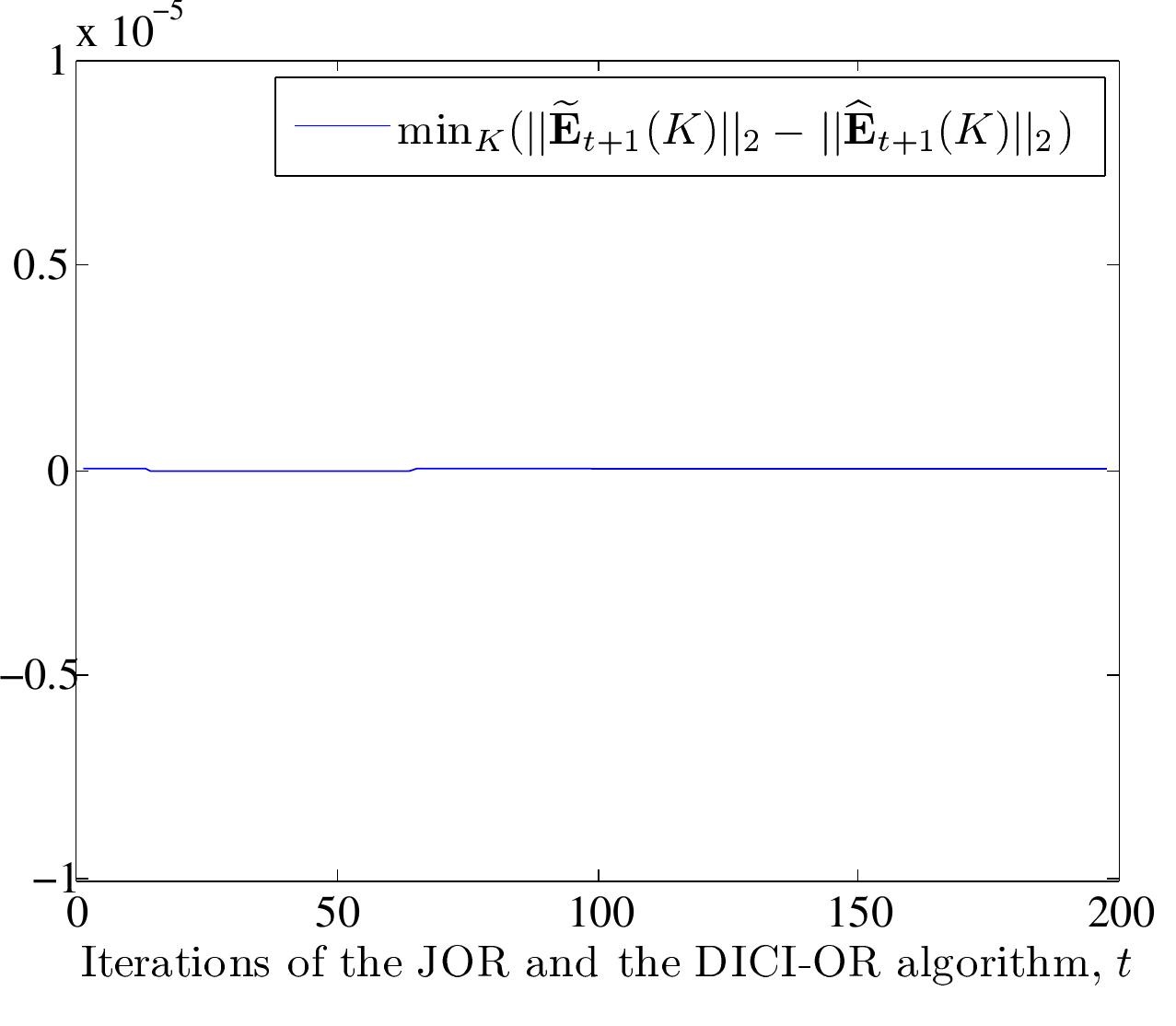}
}
\hspace{.5cm}
\subfigure[]
{
    \label{eb3.pdf}
    \includegraphics[width=1.8in]{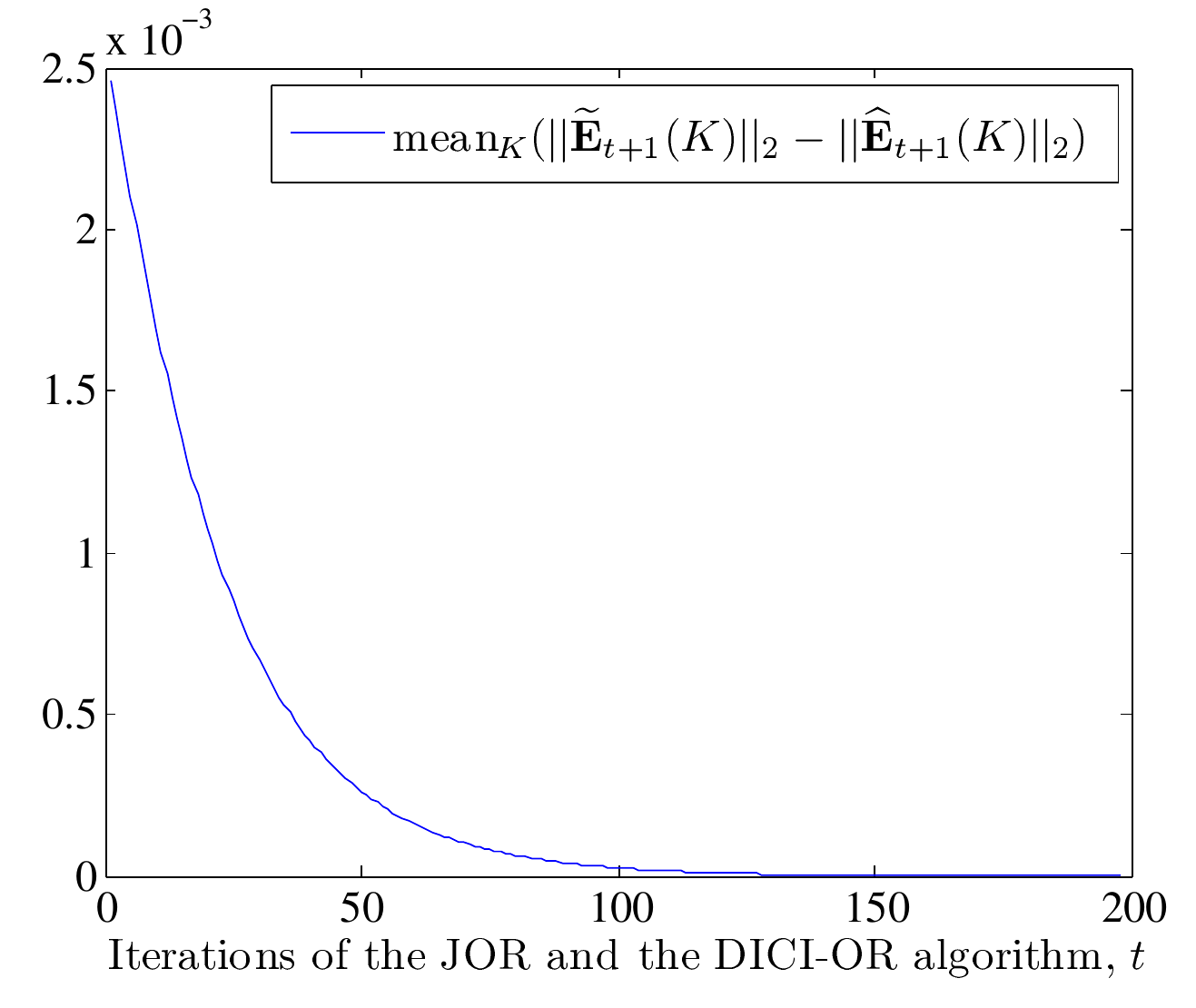}
}
\caption{Simulation for the error bound of the DICI-OR algorithm.}
\label{cntr_fig} % caption for the whole figure
\end{figure}
This verifies our claim numerically and provides us an upper bound on the spectral norm of the error process of the DICI algorithm.

\section{Local Information Filters: Initial Conditions and Local Filter Step}\label{LIF:ic_lfs}
The initial conditions and the local filter step of the LIFs are presented in the next subsections.
\subsection{Initial Conditions}
The initial condition on the local predictor is \begin{equation}\label{icz}\mathbf{\widehat{z}}_{0|-1}^{(l)} = \mathbf{0}.\end{equation} Since the local information matrix and the local error covariances are not the inverse of each other, \eqref{Zn2S}, we obtain the initial condition on the prediction information matrix by using the~$L-$banded inversion theorem~\cite{kavcic:00}, provided in appendix~\ref{LBth}. This step may require a local communication step further elaborated in section~\ref{PD_1}. \begin{equation}\label{icZ}\mathbf{Z}_{0|-1}^{(l)}\overleftarrow{\mbox{$\qquad$$L-$Banded Inversion Theorem$\qquad$}}\mathbf{S}_{0}^{(l)}\end{equation}

\subsection{Local Filter Step}
In this section, we present the local filter step of the LIFs. The local filter step is given by \begin{subequations}\label{loc_filt_step}\begin{eqnarray}\label{loc_filt_step_pt1}
\mathbf{Z}^{(l)}_{k|k} &=& \mathbf{Z}^{(l)}_{k|k-1}\mathbf{+\mathcal{I}}_{f}^{(l)},\\\label{loc_filt_step_pt2}
\mathbf{\widehat{z}}^{(l)}_{k|k}&=&\mathbf{\widehat{z}}^{(l)}_{k|k-1}\mathbf{+i}_{f,k}^{(l)},\end{eqnarray}\end{subequations} where~$\mathbf{\mathcal{I}}_{f}^{(l)}$ and $\mathbf{i}_{f,k}^{(l)}$ denote the fused observation variables. Fusion of the observations is presented in section~\ref{obs_fus}. The distribution of the addition operation, `$+$', in~\eqref{glob_filt_step} is straightforward in~\eqref{loc_filt_step}. Recall that the observation fusion,~\eqref{o_f_waa}, is carried out using the iterative weighted averaging algorithm. The asymptotic convergence of this iterative algorithm is guaranteed under certain conditions, e.g., connected sensor communication sub-graph,~$\mathcal{G}_j$, see~\cite{boyd:04}, for further details. Hence, with the required assumptions on the sub-graph,~$\mathcal{G}_j$, the observation fusion algorithm,~\eqref{o_f_waa}, asymptotically converges, and hence (with a slight abuse of notation), \begin{equation}\bigcup_{l=1}^N\mathbf{i}_{f,k}^{(l)}\rightarrow\mathbf{i}_{k}\mbox{    and    } \bigcup_{l=1}^N\mathbf{\mathcal{I}}_{f}^{(l)}\rightarrow\mathbf{\mathcal{I}}.\end{equation} The above notation implies that the local fused information variables,~$\mathbf{\mathcal{I}}_{f}^{(l)}$ and~$\mathbf{i}_{f,k}^{(l)}$, when combined over the entire sensor network, asymptotically converge to the global information variables,~$\mathbf{\mathcal{I}}$ and $\mathbf{i}_{k}$. This, in turn, implies that the local filter step of the LIFs asymptotically converges to the global filter step,~\eqref{glob_filt_step}, of the CLBIF.

Once the local filter step is completed, the DICI algorithm is employed on the local information matrices,~$\mathbf{Z}_{k|k}^{(l)}$ obtained from~\eqref{loc_filt_step_pt1}, to convert them into the local error covariance matrices,~$\mathbf{S}_{k|k}^{(l)}$. Finally, to convert the estimates in the information domain,~$\mathbf{\widehat{z}}_{k|k}^{(l)}$, to the estimates in the Kalman filter domain, $\mathbf{\widehat{x}}_{k|k}^{(l)}$, we specialize the DICI algorithm to matrix-vector product~\eqref{z2x}.

\section{Local Information Filters: Local Prediction Step}\label{LIF:lps}
This section presents the distribution of the global prediction step,~\eqref{glob_pred_step}, into the local prediction step at each LIF. This section requires the results of the DICI algorithm for the~$L-$banded matrices, introduced in section~\ref{dist_jac}.

\subsection{Computing the local prediction information matrix,~$\mathbf{Z}^{(l)}_{k|k-1}$}\label{PD_1}
Because of the coupled local dynamics of the reduced sensor-based models, each sensor may require that some of the estimated states be communicated as internal inputs,~$\mathbf{\widehat{d}}_{k|k}^{(l)}$, to its LIF, as shown in~\eqref{dist_dyn}. These states are the directed edges into each cut-point set in Figure~\ref{cutsets_fig}. Hence, the error associated to a local estimation procedure is also influenced by the error associated to the neighboring estimation procedure, from where the internal inputs are being communicated. This dependence is true for all sensors and is reflected in the local prediction error covariance matrix,~$\mathbf{S}_{k|k-1}^{(l)}$, as it is a function of the global estimation error covariance matrix,~$\mathbf{S}_{k-1|k-1}$. Equation~\eqref{Pikk1l} follows from \eqref{g_pred_pt1} after expanding~\eqref{g_pred_pt1} for each diagonal submatrix,~$\mathbf{S}_{k|k-1}^{(l)}$, in~$\mathbf{S}_{k|k}$. \begin{equation}\label{Pikk1l}\mathbf{S}_{k|k-1}^{(l)}=\mathbf{F}_{l}\mathbf{S}_{k-1|k-1}\mathbf{F}_{l}^{T}+\mathbf{G}^{(l)}\mathbf{Q}^{(l)} \mathbf{G}^{(l)T}.\end{equation} The matrix,~$\mathbf{F}_l=\mathbf{T}_l\mathbf{F}$, (the matrix $\mathbf{T}_l$ is introduced in~\eqref{Tl}) is an~$n_l\times n$ matrix, which relates the state vector,~$\mathbf{x}_k$, to the local state vector,~$\mathbf{x}_k^{(l)}$. Figure~\ref{Fl_fig} shows that the matrix,~$\mathbf{F}_l$, is further divided into~$\mathbf{F}^{(l)}$ and~$\mathbf{D}^{(l)}$. With this sub-division of $\mathbf{F}_l$, the first term on the right hand side of~\eqref{Pikk1l},~$\mathbf{F}_{l}\mathbf{S}_{k-1|k-1}\mathbf{F}_{l}^{T}$, can be expanded, and \eqref{Pikk1l} can be written as \begin{eqnarray}\label{Pikk1l_t}\mathbf{S}_{k|k-1}^{(l)}&=&\mathbf{F}^{(l)}\mathbf{S}_{k-1|k-1}^{(l)}\mathbf{F}^{(l)T}+
\mathbf{F}^{(l)}\mathbf{S}_{k-1|k-1}^{x^{(l)}d^{(l)}}\mathbf{D}^{(l)T}\nonumber\\&+&\left(\mathbf{F}^{(l)}\mathbf{S}_{k-1|k-1}^{x^{(l)}d^{(l)}}\mathbf{D}^{(l)T}\right)^T+ \mathbf{D}^{(l)}\mathbf{S}_{k-1|k-1}^{d^{(l)}d^{(l)}}\mathbf{D}^{(l)T}+\mathbf{G}^{(l)}\mathbf{Q}^{(l)} \mathbf{G}^{(l)T},\end{eqnarray} where

$\mathbf{S}_{k-1|k-1}^{(l)}$ is the local error covariance matrix, which is available from~\eqref{loc_filt_step_pt1} and the DICI algorithm at sensor~$l$;

$\mathbf{S}_{k-1|k-1}^{d^{(l)}d^{(l)}}$ is the local error covariance matrix, which is available from~\eqref{loc_filt_step_pt1} and the DICI algorithm at the sensors having the states,~$\mathbf{d}_k^{(l)}$, in their reduced models;

$\mathbf{S}_{k-1|k-1}^{x^{(l)}d^{(l)}}$ is the error cross correlation between the local state vector,~$\mathbf{x}_k^{(l)}$, and the local internal input vector,~$\mathbf{d}_k^{(l)}$.

The non~$L-$banded entries in this matrix can be computed from the equation~\eqref{s35_non}, see~\cite{asif:05}. Since the model matrix,~$\mathbf{F}$, is sparse, we do not need the entire error covariance matrix,~$\mathbf{S}_{k-1|k-1}$, only certain of its submatrices. Since the model matrix,~$\mathbf{F}$, is localized, long-distance communication is not required, and the submatrices are available at the neighboring sensors.

Once we have calculated the local prediction error covariance matrix,~$\mathbf{S}_{k|k-1}^{(l)}$, we realize~\eqref{Zn2S} and compute the local prediction information matrix, $\mathbf{Z}_{k|k-1}^{(l)}$, using the~$L-$banded Inversion Theorem (see~\cite{kavcic:00} and appendix~\ref{LBth}).\begin{equation}\label{Skk12Zkk1}\mathbf{Z}_{k|k-1}^{(l)}\overleftarrow{\mbox{$\qquad L-$Banded Inversion Theorem$\qquad$}}\mathbf{S}_{k|k-1}^{(l)}.\end{equation} From~\eqref{L_bnd_inv} in appendix~\ref{LBth}, to calculate the local prediction information matrix,~$\mathbf{Z}_{k|k-1}^{(l)}$, we only need the~$\mathbf{S}_{k|k-1}^{(l)}$ from sensor `$l$' and from some additional neighboring sensors. Hence $\mathbf{Z}_{k|k-1}^{(l)}$ is again computed with only local communication and~$n_l$th order computation.

\subsection{Computing the local predictor,~$\mathbf{\widehat{z}}^{(l)}_{k|k-1}$}
We illustrate the computation of the local predictor,~$\mathbf{\widehat{z}}^{(3)}_{k|k-1}$, for the $5-$dimensional system,~\eqref{ex_sys1}$-$\eqref{ex_sys2}, with~$L=1$. The local predictor,~$\mathbf{\widehat{z}}^{(3)}_{k|k-1}$, at sensor~$3$ follows from the global predictor,~\eqref{g_pred_pt2}, and is given by \begin{equation}\label{zkk11}
\mathbf{\widehat{z}}^{(3)}_{k|k-1} = \mathbf{Z}_{k|k-1}^{(3)}\left(\mathbf{F}^{(3)}\mathbf{\widehat{x}}^{(3)}_{k-1|k-1}+\mathbf{D}^{(3)}\mathbf{\widehat{d}}^{(3)}_{k-1|k-1}\right)+
\left[\begin{array}{c}
z_{34}\left(f_{31}\widehat{x}_{1,k-1|k-1}^{(1)}+f_{33}\widehat{x}_{3,k-1|k-1}^{(2)}\right)\\
0
\end{array}\right],\end{equation}
%\mathbf{\widehat{z}}^{(1)}_{k|k-1} = \mathbf{Z}_{k|k-1}^{(1)}\left(\mathbf{F}^{(1)}\mathbf{\widehat{x}}^{(1)}_{k-1|k-1}+\mathbf{D}^{(1)}\mathbf{\widehat{d}}^{(1)}_{k-1|k-1}\right)+
%\left[\begin{array}{c}
%0 \\
%0\\
%z_{34}\left(f_{43}\widehat{x}_{3,k-1|k-1}^{(2)}+f_{45}\widehat{x}_{5,k-1|k-1}^{(3)}\right)
%\end{array}\right],
where~$z_{34}$ is the only term arising due to the~$L=1-$banded (tridiagonal) assumption on the prediction information matrix, $\mathbf{Z}_{k|k-1}$. Note that~$f_{31}\widehat{x}_{1,k-1|k-1}^{(1)}+f_{33}\widehat{x}_{3,k-1|k-1}^{(2)}$ is a result of~$\mathbf{f}_3\mathbf{\widehat{x}}_{k-1|k-1}$, where~$\mathbf{f}_3$ is the third row of the model matrix,~$\mathbf{F}$. A model matrix with a localized and sparse structure ensures that $\mathbf{f}_3\mathbf{\widehat{x}}_{k-1|k-1}$ is computed from a small subset of the estimated state vector,~$\mathbf{\widehat{x}}_{k-1|k-1}^{(\mathcal{Q})}$, communicated by a subset~$\mathcal{Q}\subseteq\mathcal{K}(l)$ of the neighboring sensors, which are modeling these states in their reduced models. This may require multi-hop communication.

Generalizing, the local predictor in the information domain,~$\mathbf{\widehat{z}}_{k|k-1}^{(l)}$, is given by
\begin{eqnarray}\label{loc_z_comp_t}\mathbf{\widehat{z}}^{(l)}_{k|k-1} &=& \mathbf{Z}_{k|k-1}^{(l)}\left(\mathbf{F}^{(l)}\mathbf{\widehat{x}}^{(l)}_{k-1|k-1}+\mathbf{D}^{(l)}\mathbf{\widehat{d}}^{(l)}_{k-1|k-1}\right)+
f_1\left(\mathbf{Z}_{k|k-1}^{(\mathcal{V})},\mathbf{F}^{(\mathcal{V})},\mathbf{\widehat{x}}^{(\mathcal{Q})}_{k-1|k-1}\right)\end{eqnarray} for some~$\mathcal{V,Q} \subseteq \mathcal{K}(l)$, where~$f_1(\cdot)$ is a linear function and depends on~$L$.

\subsection{Estimate Fusion}\label{est_fus}
We present the following fact.

{\bf Fact:} Let $m$ denote the iterations of the consensus algorithm that is employed to fuse the observations. As $m\rightarrow\infty$, the local estimates, $\mathbf{\widehat{z}}_{k|k}^{(l)}$, in~\eqref{loc_filt_step_pt2} also reach a consensus on the estimates of the shared states.

It is straightforward to note that as $m\rightarrow\infty$ we have a consensus on the estimates (of the shared states) in the local filter step~\eqref{loc_filt_step_pt2}, if we have a consensus on the local predictors (of the shared states), $\mathbf{\widehat{z}}_{k|k-1}$. To show the consensus on the local predictors, we refer back to our illustration and write the local predictors for sensor $2$ as follows,\begin{eqnarray}\label{one}\mathbf{\widehat{z}}^{(2)}_{k|k-1} &=& \mathbf{Z}_{k|k-1}^{(2)}\left(\mathbf{F}^{(2)}\mathbf{\widehat{x}}^{(2)}_{k-1|k-1}+\mathbf{D}^{(2)}\mathbf{\widehat{d}}^{(2)}_{k-1|k-1}\right)+
\left[\begin{array}{c}
z_{12}\left(f_{11}\widehat{x}_{1,k-1|k-1}^{(1)}+f_{12}\widehat{x}_{2,k-1|k-1}^{(2)}\right) \\
0\\
z_{45}\left(f_{54}\widehat{x}_{4,k-1|k-1}^{(2)}+f_{55}\widehat{x}_{5,k-1|k-1}^{(3)}\right)
\end{array}\right].\end{eqnarray} The predictor for the shared state $x_{4,k}$ can now be extracted from \eqref{zkk11} and \eqref{one} and can be verified to be the following for $l=2,3$.\begin{equation}{\widehat{z}}^{(l)}_{4,k|k-1} = z_{34} f_{31}{\widehat{x}}_{1,k-1|k-1}^{(1)}+(z_{34}f_{33}+z_{44}f_{43}){\widehat{x}}_{3,k-1|k-1}^{(2)}+z_{45}f_{54}{\widehat{x}}_{4,k-1|k-1}^{(3)} +(z_{44}f_{45}+z_{45}f_{55}){\widehat{x}}_{5,k-1|k-1}^{(3)}\end{equation}
The elements $z_{ij}$ belong to the prediction information matrix, which is computed using the DICI algorithm and the $L-$banded inversion theorem. It is noteworthy that the DICI algorithm is not a consensus algorithm and thus the elements $z_{ij}$ are the same across the sensor network at any iteration of the DICI algorithm. With a consensus on the local predictors, the iterations of the consensus algorithm on the observations lead to a consensus on the shared estimates.
%
%For each state~$x_j$, the LIF at each sensor~$s\in \mathcal{G}_j$ provides an estimate of this state,~$\widehat{x}_{j,k|k}^{(s)}$. If~$\mathcal{G}_j$ contains more than one sensor, there are multiple correlated estimates of the same state, which should be fused in order to obtain an estimate with smaller variance, a problem considered by Durbin~\cite{durbin:59}; its vector extensions can be found in~\cite{schweppe_book,willsky:81,willsky:82}. At each sensor~$s\in \mathcal{G}_j$, let $\pi^{(s)}_j$ be the variance of the~$j$th state estimate,~$\widehat{x}_{j,k|k}^{(s)}$, where~$\pi^{(s)}_j$ is a diagonal element in the local estimation error covariance matrix,~$\mathbf{S}^{(s)}_{k|k}$, at sensor~$s$. We fuse the estimates using the parallel fusion of estimates formula, which for scalar quantities can be derived by using Lagrange multipliers and is given by \begin{equation}\label{ef}\widehat{x}_{j,k|k}=\left(\sum_{s\in \mathcal{G}_j} \left(\pi^{(s)}_j\right)^{-1} \right)^{-1}\left(\sum_{s\in \mathcal{G}_j}\left(\pi^{(s)}_j\right)^{-1}\widehat{x}_{j,k|k}^{(s)}\right).\end{equation} Each of the sums in~\eqref{ef} is carried out using the weighted averaging algorithm~\cite{boyd:04}, and then combined to compute~\eqref{ef} at each sensor~$s\in \mathcal{G}_j$.

\section{Results}\label{res}
\subsection{Summary of the LIFs}
We summarize the distributed local Information filters. The initial conditions are given by~\eqref{icz} and~\eqref{icZ}. Observation fusion is carried out using \eqref{o_f_waa}. The fused observation variables,~$\mathbf{i}^{(l)}_{f,k}$ and~$\mathbf{I}^{(l)}_{f,k}$, are then employed in the local filter step, \eqref{loc_filt_step_pt1} and~\eqref{loc_filt_step_pt2}, to obtain the local information matrix and the local estimator,~$\mathbf{\widehat{Z}}_{k|k}^{(l)}$ and $\mathbf{z}_{k|k}^{(l)}$, respectively. We then implement the DICI algorithm~\eqref{dici1}$-$\eqref{dici2} and~\eqref{sij_non} to compute the local error covariance matrix,~$\mathbf{S}_{k|k}^{(l)}$, from the local information matrix,~$\mathbf{Z}_{k|k}^{(l)}$. The DICI algorithm is again employed to compute the local estimates in the Kalman filter domain,~$\mathbf{\widehat{x}}_{k|k}^{(l)}$, from the local estimator,~$\mathbf{\widehat{z}}_{k|k}^{(l)}$, as a special case. Finally the local prediction step is completed by computing the local prediction error covariance matrix,~$\mathbf{\widehat{S}}_{k|k-1}^{(l)}$, the local prediction information matrix, $\mathbf{\widehat{Z}}_{k|k-1}^{(l)}$, and, the local predictor,~$\mathbf{\widehat{z}}_{k|k-1}^{(l)}$, from~\eqref{Pikk1l_t},~\eqref{Skk12Zkk1}, and~\eqref{loc_z_comp_t}, respectively.

\subsection{Simulations}
\begin{figure}
\centering
\subfigure[]
{
    \label{F_sim_fig123.pdf}
    \includegraphics[height=2.1in]{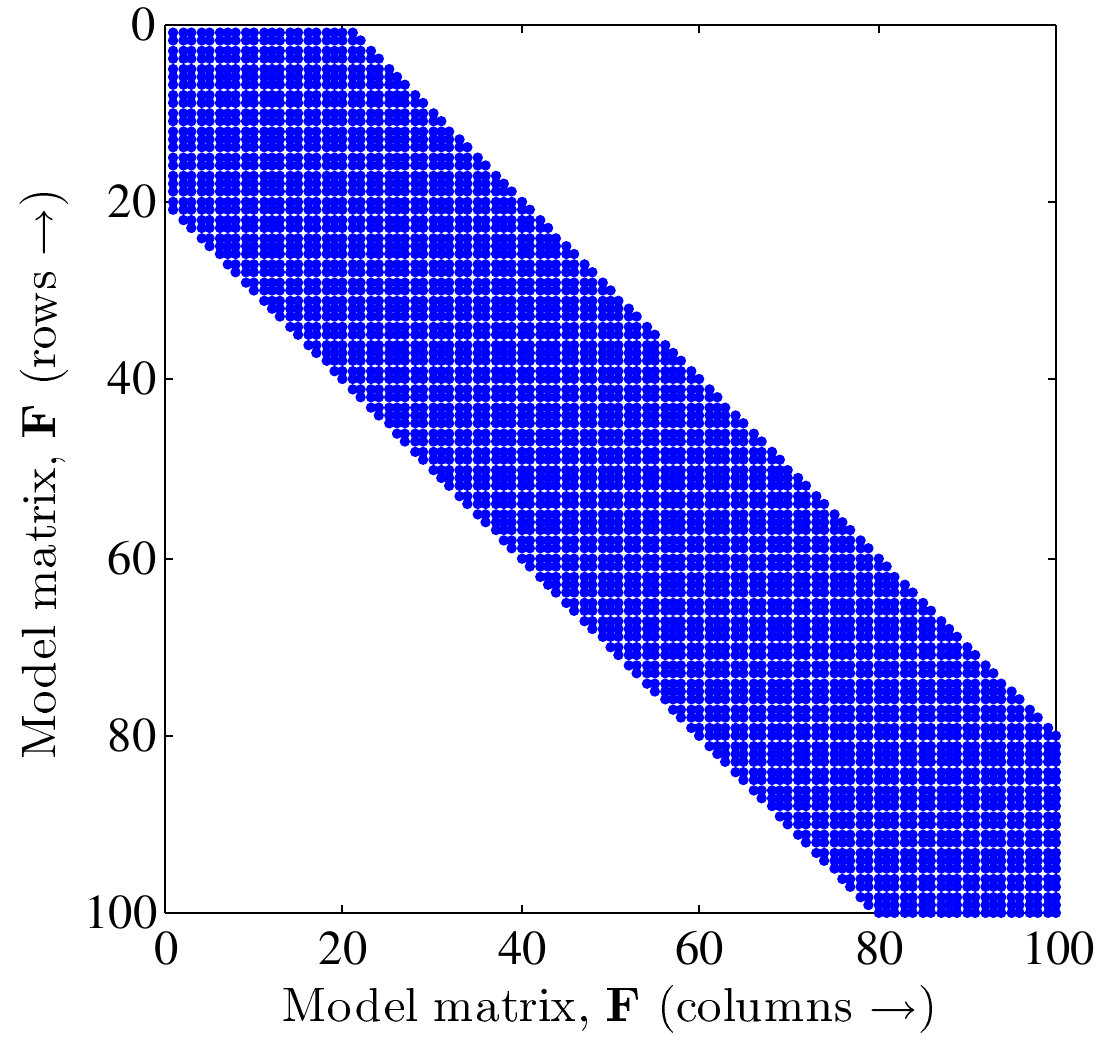}
}
\hspace{1cm}
\subfigure[]
{
    \label{F_sim_fig4.pdf}
    \includegraphics[height=2.1in]{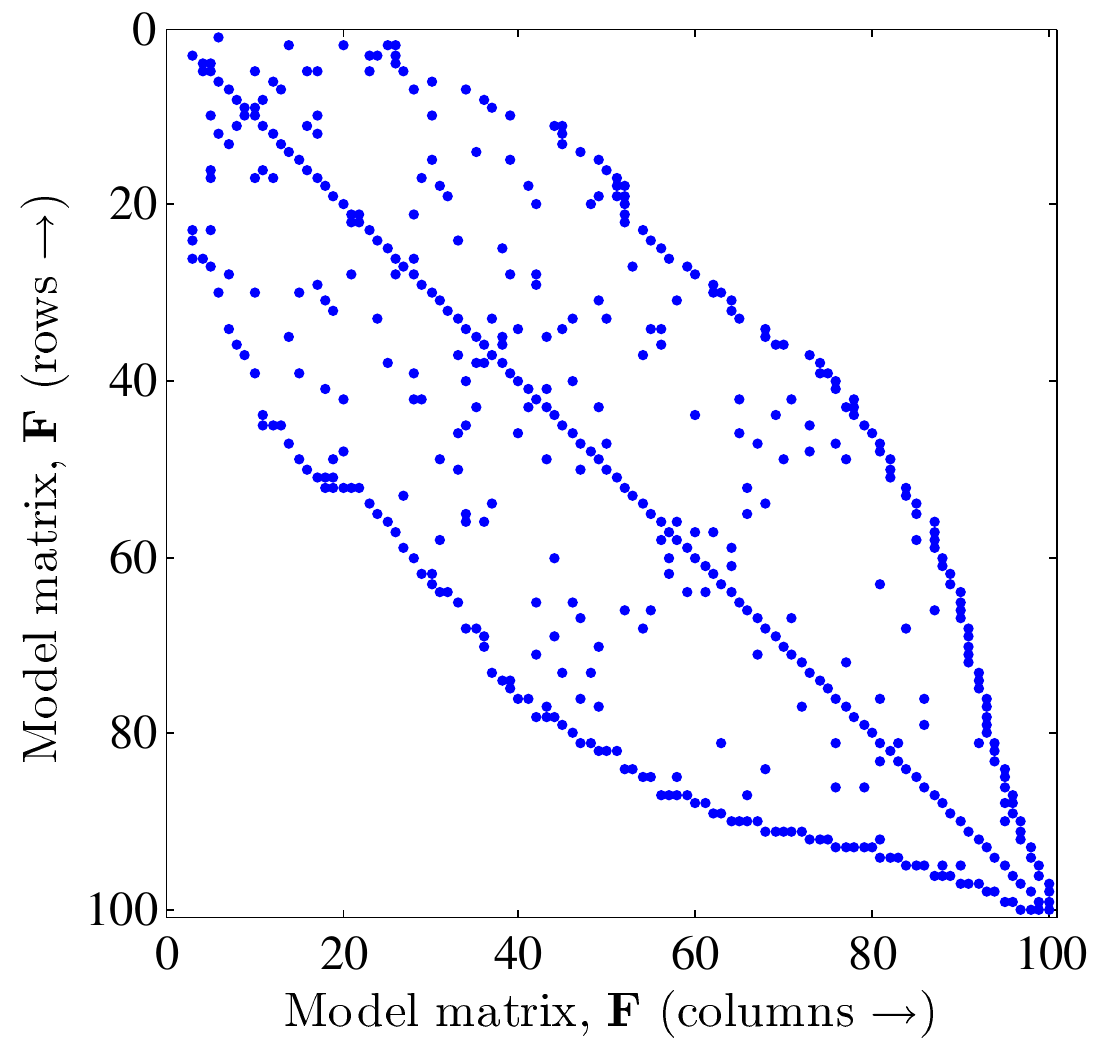}
}
\hspace{1cm}
\subfigure[]
{
    \label{sim_fig3.pdf}
    \includegraphics[height=2.1in]{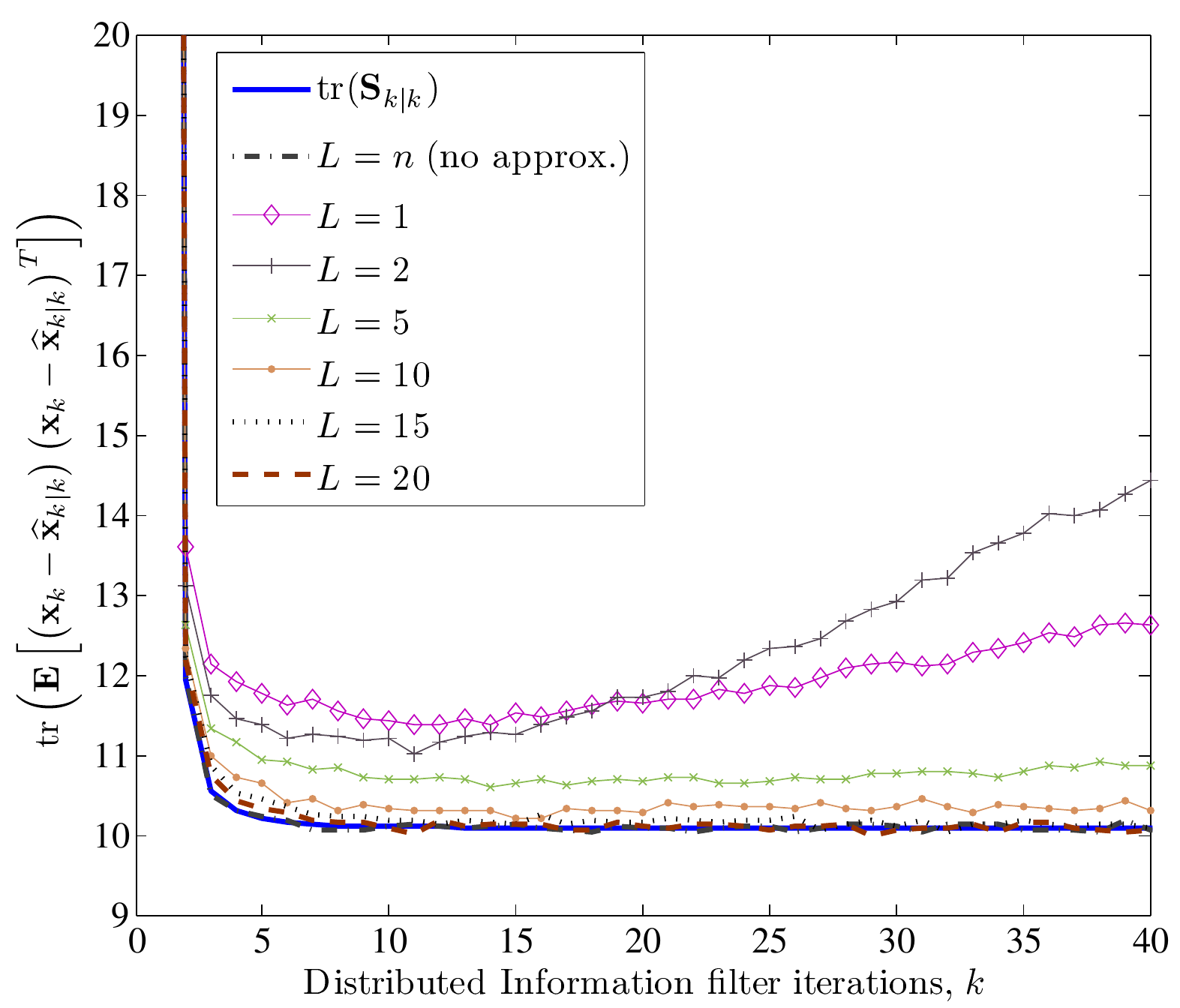}
}
\subfigure[]
{
    \label{sim_fig4.pdf}
    \includegraphics[height=2.1in]{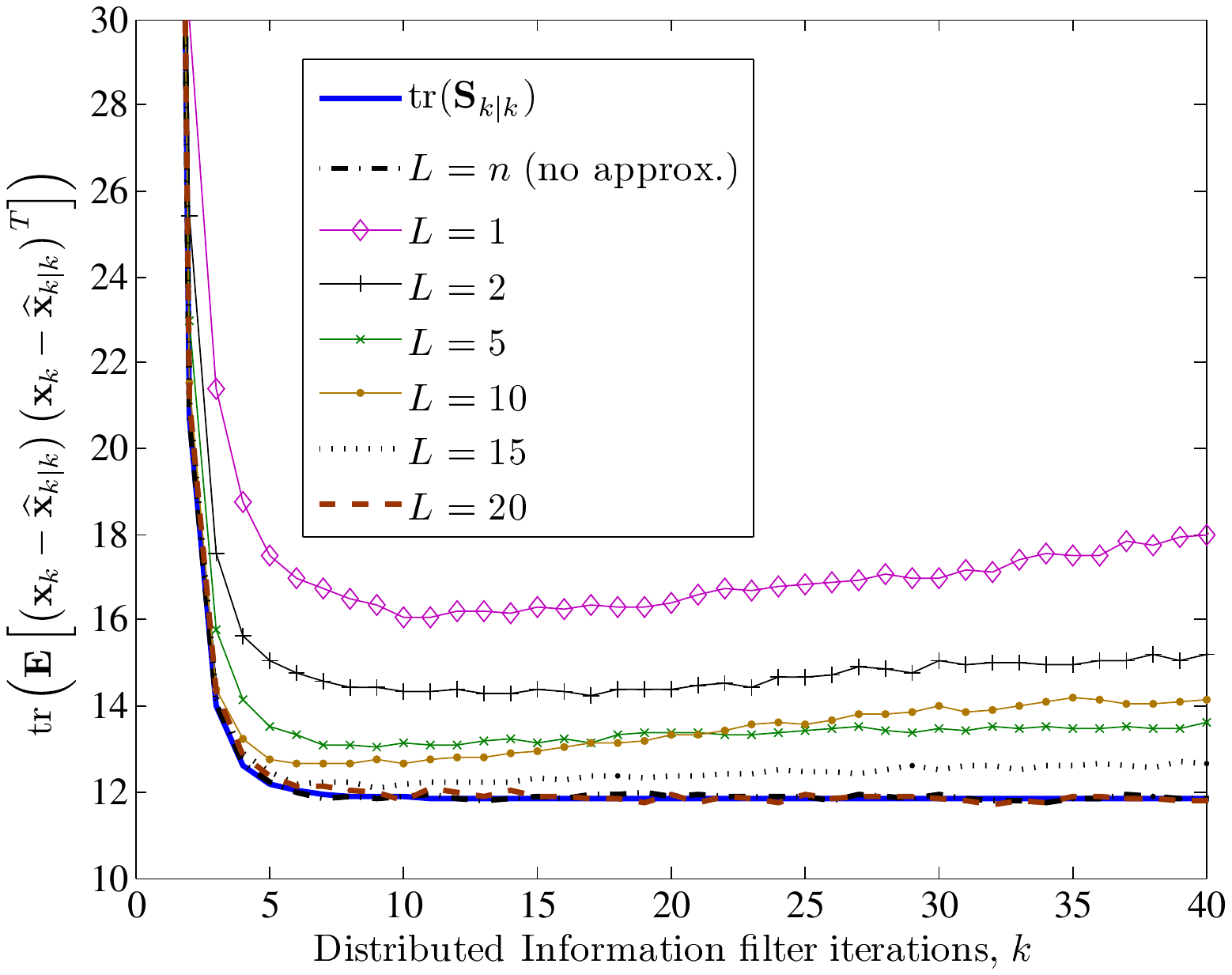}
}
\caption{(a \& b) Non-zero elements (chosen at random) of $100\times 100$, $L=20-$banded (Figure~\ref{F_sim_fig123}) and $L=36-$banded (Figure~\ref{F_sim_fig4}) model matrices, $\mathbf{F}$, such that $||\mathbf{F}||_2=1$. (c \& d) Distributed Kalman filter is implemented on the model matrices in Figure~\ref{F_sim_fig123}-\ref{F_sim_fig4} and the global observation matrix, $\mathbf{H}$ (Figure~\ref{H_fig}), in Figure~\ref{sim_fig3}-\ref{sim_fig4}. The expectation operator in the trace (on horizontal axis) is simulated over $1000$ Monte Carlo trials.}
\label{}
\end{figure}

\begin{figure}
\centering
\includegraphics[width=5in]{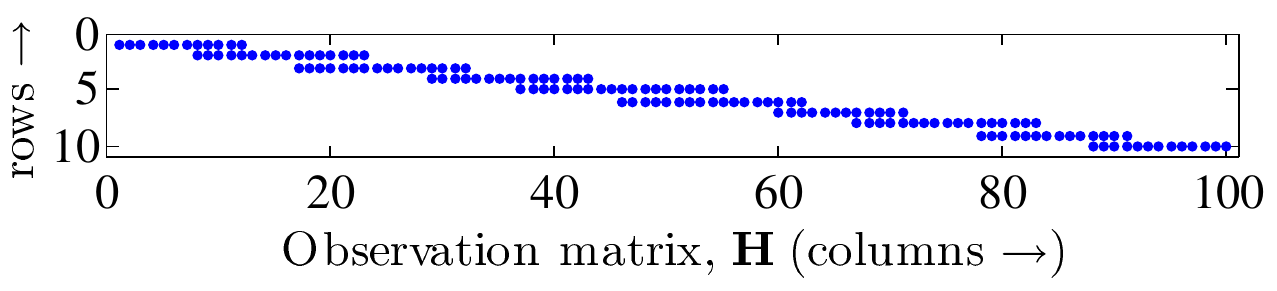}
\caption{Global observation matrix, $\mathbf{H}$. The non-zero elements (chosen at random) are shown. There are $N=10$ sensors, where the $l$th row of $\mathbf{H}$ corresponds to the local observation matrix, $\mathbf{H}_l$, at sensor $l$. The overlapping states (for which fusion is required) can be seen as the overlapping portion of the rows.}
\label{H_fig}
\end{figure}

We simulate a~$n=100-$dimensional system with $N=10$ sensors monitoring the system. Figures~\ref{F_sim_fig123} and~\ref{F_sim_fig4} show the non-zero elements (chosen at random) of the model matrix, $\mathbf{F}$, such that $||\mathbf{F}||_2=1$. The model matrix in Figure~\ref{F_sim_fig123} is $L=20-$banded. The model matrix in Figure~\ref{F_sim_fig4} is $L=36-$banded that is obtained by employing the reverse Cuthill-Mckee algorithm \cite{mckee:69} for bandwidth reduction of a sparse random $\mathbf{F}$. The non-zeros (chosen at random as $\mbox{Normal}(0,1)$) of the global observation matrix, $\mathbf{H}$, are shown in Figure~\ref{H_fig}. The $l$th row of the global observation matrix, $\mathbf{H}$, is the local observation matrix, $\mathbf{H}_l$, at sensor $l$. Distributed Kalman filters are implemented on (i) $\mathbf{F}$ in \ref{F_sim_fig123} and $\mathbf{H}$ in Figure~\ref{H_fig}; and (ii) $\mathbf{F}$ in \ref{F_sim_fig4} and $\mathbf{H}$ in Figure~\ref{H_fig}. The trace of the error covariance matrix, $\mathbf{S}_{k|k}$, is simulated for different values of $L$ in $[n,1,2,5,10,15,20]$ and the plots are shown (after averaging over $1000$ Monte Carlo trials) in Figure~\ref{sim_fig3} for case (i); and in Figure~\ref{sim_fig4} for case (ii). The stopping criteria for the DICI algorithm and the consensus algorithm are such that the deviation in their last $10$ iterations is less than $10^{-5}$. In both Figure~\ref{sim_fig3} and Figure~\ref{sim_fig4}, $\mbox{tr}(S_{k|k})$ represents the trace of the solution of the Riccati equation in the CIF (no approximation).

\begin{figure}
\centering
\includegraphics[width=2.5in]{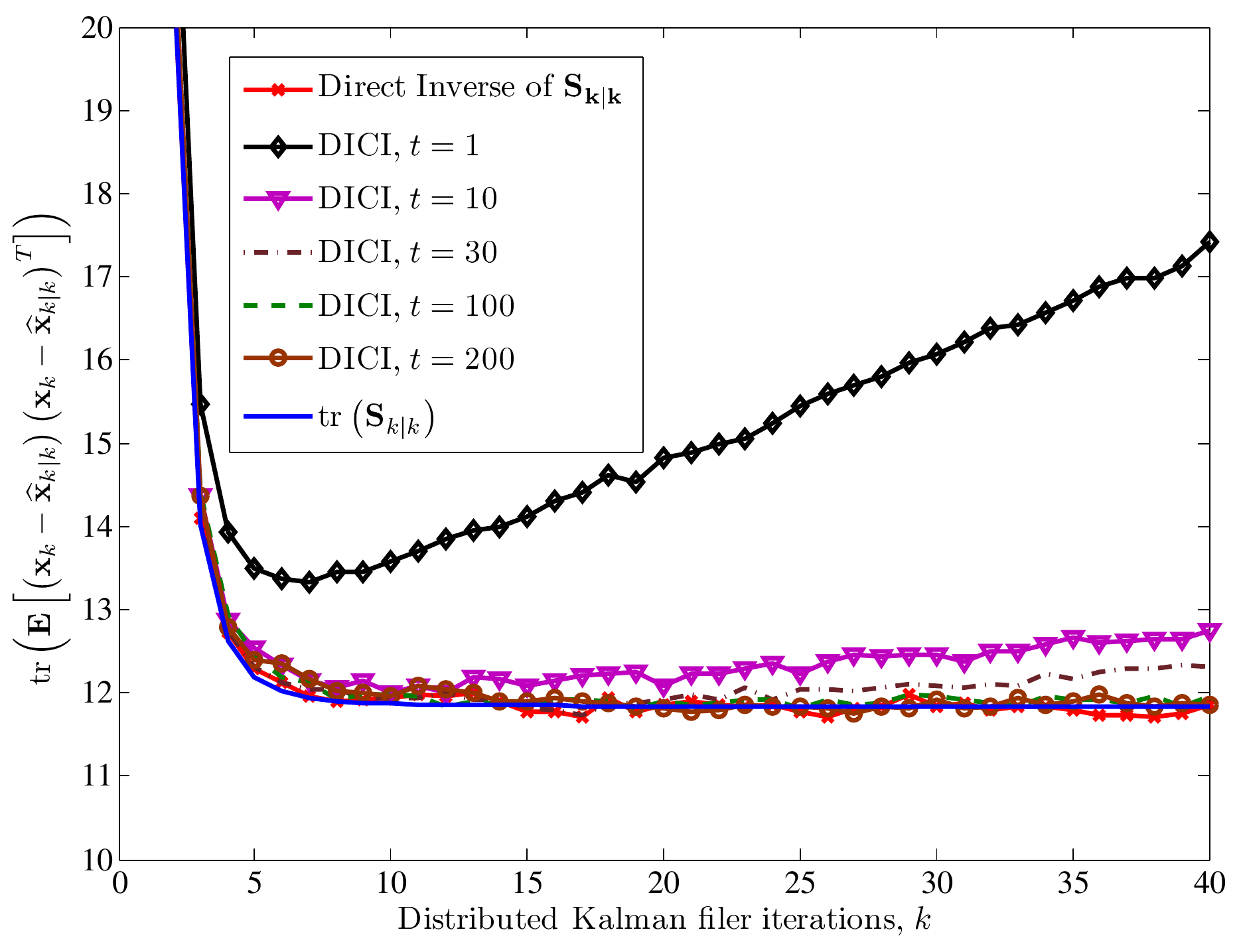}
\caption{Performance of the DICI algorithm as a function of the  number of DICI iterations, $t$.}
\label{sim_DICI_fn1}
\end{figure}

With $1000$ Monte Carlo trials, we further simulate the trace of the error covariance, $\mbox{tr}(S_{k|k})$, for case (ii) and $L=20-$banded approximations as a function of the number of iterations, $t$, of the DICI-OR algorithm. We compare this with (a) the simulation obtained from the $O(n^3)$ direct inverse of the the error covariance (with $L=20-$banded approximation on its inverse); and (b) $\mbox{tr}\left(\mathbf{S}_{k|k}\right)$, trace of the solution of the Riccati equation of the CIF (no approximation). We choose $t=[1,10,30,100,200]$ for the DICI algorithm and show the results in Figure~\ref{sim_DICI_fn1}. As $t\uparrow$, the curves we obtain from the DICI algorithm get closer to the curve we obtain with the direct inverse.

The simulations confirm the following: (i) The LIFs asymptotically track the results of the CLBIF, see Figure~\ref{sim_DICI_fn1}; (ii) We verify that as~$L\uparrow$, the performance is virtually indistinguishable from that of the CIF, as pointed out in~\cite{asif:05}; this is in agreement with the fact that the approximation is optimal in Kullback-Leibler sense, as shown in~\cite{kavcic:00}. Here, we also point out that, as we increase $L$ the performance increases, but, we pay a price in terms of the communication cost, as we may have to communicate in a larger neighborhood. (iii) If we increase the number of Monte Carlo trials the variations reduce, and the filters eventually follow the solution of the Riccati equation,~$\mbox{tr}\left(\mathbf{S}_{k|k}\right)$; (iv) The curve in Figure~\ref{sim_DICI_fn1} with $t=1$ shows the decoupled LIFs, when the global error covariance is treated as a block diagonal matrix. This is unstable as we do not fuse the error covariances and the individual sub-systems are not observable. We next discuss the computational advantage of the LIFs over some of the existing methods.

\subsection{Complexity}
We regard the multiplication of two~$n\times n$ matrices as an~$O(n^3)$ operation, inversion of an~$n\times n$ matrix also as an~$O(n^3)$ operation and multiplication of an~$n\times n$ matrix and an~$n\times 1$ vector as an~$O(n^2)$ operation. For all of the following, we assume~$N$ sensors monitoring the global system, \eqref{glob_dyn}.

\subsubsection{Centralized Information Filter, CIF}\label{CIF_comp}
This is the case where each sensor sends its local observation vector to a centralized location or a fusion center, where the global observation vector is then put together. The fusion center then implements the CIF, with an~$O(n^3)$ computation complexity for each time~$k$, so we have the complexity as~$O(n^3k)$, with inordinate communication requirements (back and forth communication between the sensors and the central location).

\subsubsection{Information Filters with Replicated Models at each Sensor and Observation Fusion}\label{DIF_comp}
In this scheme, the local observation vectors are not communicated to the central location, but are fused over an all-to-all communication network, \cite{raowhyte:91}, or an arbitrary network,~\cite{olfati:05}, that requires an iterative algorithm. Computational complexity at each sensor is~$O(n^3k)$ in~\cite{raowhyte:91}. In~\cite{olfati:05}, let the complexity of the iterative algorithm be~$O(\varphi(n))$ and let $t_\varphi$ be the number of iterations required by the iterative algorithm to converge. Each sensor implements a CIF after fusing the observations. For each time index,~$k$, each sensor requires~$O(n^3)$ operations plus the operations required for the iterative algorithms, which are~$O(\varphi(n)t_\varphi)$; so the total computation complexity is $O((n^3+\varphi(n)t_\varphi)k)$ at each sensor. Communication requirements are global in~\cite{raowhyte:91}, and local in~\cite{olfati:05}.

\subsubsection{Distributed Kalman Filters: Local Information Filters, LIFs}
The distributed Kalman filter presented in this paper has three iterative algorithms. In all the other steps the computation complexity is dominated by~$O(n_l^3)$, where~$n_l \ll n$. Let~$t_o$ be the iterations required by the weighted averaging algorithm, where at each step of the iterative algorithm the computations are dominated by~$O(n_l^2)$. Let~$t_{J_1}$ be the iterations required by the DICI algorithm for vectors, where at each step of the iterative algorithm the computations are dominated by an~$O(L^2)$ operations. Let~$t_{J_2}$ be the iterations required by the DICI algorithm, where at each step of the iterative algorithm the computations are dominated by~$O(L^4)$ operations. Recalling that $L\sim n_l$ from section~\ref{mod_dist}, the total computation complexity is $O((n_l^3+n_l^2t_o+n_l^2t_{J_1}+n_l^4t_{J_2})k)$. Let~$t_{\varpi}=\mbox{max}(t_o,t_{J_1},t_{J_2})$, then the computation complexity is bounded by~$O(n_l^4t_\varpi k)$ at each sensor for the LIFs, which is much smaller than the computational cost of the solutions in~\ref{CIF_comp} and~\ref{DIF_comp}. The communication requirement in the LIFs may be multi-hop but is always constrained in a neighborhood because of the structural assumptions on the model matrix, $\mathbf{F}$.

\section{Conclusions}\label{conc}
In conclusion, we presented a \emph{distributed} implementation of the Kalman filter for sparse large-scale systems monitored by sensor networks. In our solution, the communication, computing, and storage is local and distributed across the sensor network, no single sensor processes~$n-$dimensional vectors or matrices, where~$n$, usually a large number, is the dimension of the state vector representing the random field. We achieve this by solving three linked problems: (1)~\emph{Distributing} the global state variable model representing the random field into sensor based reduced-order models. These reduced-order models are obtained using a graph-theoretic model distribution technique that decomposes the random field model into coupled low-order models; (2)~\emph{Fusing}, through distributed averaging, multiple sensor observations of the state variables that are common across sub-systems; and (3)~\emph{Inverting} full~$n-$dimensional matrices, with local communication only, by deriving an iterative distributed iterate collapse inversion~(DICI) algorithm. The DICI algorithm only requires matrices and vectors of the order~$n_{l}$ of the reduced state vectors. The DICI algorithm preserves the coupling among the local Kalman filters. Our solution contrasts with existing Kalman filter solutions for sensor networks that either replicate an~$n-$dimensional Kalman filter at each sensor or reduce the model dimension at the expense of decoupling the field dynamics into lower-dimensional models. The former are infeasible for large-scale systems and the latter are not optimal and further cannot guarantee any structure of the centralized error covariances..

Simulations show that the distributed implementation with local Kalman filters implemented at each sensor converges to the global Kalman filter as the number of bands in the information matrices is increased.

\appendices
\section{$L-$Banded Inversion Theorem}\label{LBth}
Let~$\mathbf{Z} = \mathbf S^{-1}$ be~$L-$banded. We apply the algorithm, given in~\cite{kavcic:00}, to obtain~$\mathbf Z$ from the submatrices in the~$L-$band of $\mathbf S$, as . We use the following notation, in~\eqref{L_bnd_not}, to partition matrix addition
and subtraction in terms of its constituent submatrices. Also~$\mathbf S_{j}^{i}$ represents the principal submatrix of~$\mathbf S$ spanning rows~$i$ through~$j$, and columns~$i$ through~$j$.

\begin{equation}\label{L_bnd_not}
\left[
\begin{array}{ccc}
a_{1} & a_{2} & 0 \\
a_{3} & x+y+z & a_{4} \\
0 & a_{5} & a_{6}%
\end{array}%
\right] = \left[
\mbox{ \unitlength1.1em
\begin{picture}(9.4,2.5)(0,9.5)
\thinlines
\footnotesize
\put(0.2,9){\framebox(2.8,2.8){$\begin{array}{ccc}a_{1} & a_{2}\\a_{3} & x \\\end{array}$}}
\put(3.5,9.8){\makebox(0,0){$+$}}
\footnotesize
\put(4,9){\framebox(1.5,1.5){$y$}}
\footnotesize
\put(6.3,9.8){\makebox(0,0){$+$}}
\footnotesize
\put(6.8,7.7){\framebox(2.8,2.8){$\begin{array}{ccc}z & a_{4}\\a_{5} & a_{5} \\\end{array}$}}
\end{picture}
}
\right]
\end{equation}

The inverse of~$\mathbf S$, when~$\mathbf{Z=S}^{-1}$ is~$L-$banded, is given by~\eqref{L_bnd_inv}, taken from~\cite{kavcic:00}, in terms of the~$L-$banded submatrices of~$\mathbf S$. Note that, to compute a principal submatrix in~$\mathbf{Z}$, we do not need the entire~$\mathbf{S}$, or even all the~$L-$banded submatrices in $\mathbf{S}$. Instead, we only require three neighboring submatrices in the~$L-$band of~$\mathbf{S}$. For proofs and further details, the interested reader can refer to~\cite{kavcic:00}.

\begin{equation}\label{L_bnd_inv}
{\bf Z} =
\left[
\mbox{ \unitlength1.1em
\begin{picture}(29.5,6)(0,6)
\thinlines
\footnotesize
\put(1,8){\framebox(3,3){${\bf S}_{L+1}^{1}$}}
\scriptsize \put(4.5,11.25){\makebox(0,0){$-1$}} \footnotesize
\put(4.75,9){\makebox(0,0){$-$}}
\footnotesize
\put(5.5,8){\framebox(2,2){${\bf S}_{L+1}^{2}$}}
\scriptsize \put(8,10.25){\makebox(0,0){$-1$}} \footnotesize
\footnotesize
\put(8.25,9){\makebox(0,0){$+$}}
\footnotesize
\put(9,7){\framebox(3,3){${\bf S}_{L+2}^{2}$}}
\scriptsize \put(12.5,10.25){\makebox(0,0){$-1$}} \footnotesize
\put(12.75,8){\makebox(0,0){$-$}}
\put(13.75,7){\makebox(0,0){$\cdot$}}
\put(14.75,6){\makebox(0,0){$\cdot$}}
\put(15.75,5){\makebox(0,0){$\cdot$}}
\put(16.75,4){\makebox(0,0){$+$}}
\footnotesize
\put(17.5,2){\framebox(3,3){${\bf S}_{N-1}^{N-L-1}$}}
\scriptsize \put(21,5.25){\makebox(0,0){$-1$}} \footnotesize
\put(21.25,3){\makebox(0,0){$-$}}
\footnotesize
\put(22,2){\framebox(2,2){${\bf S}_{N-1}^{N-L}$}}
\scriptsize \put(24.5,4.25){\makebox(0,0){$-1$}} \footnotesize
\footnotesize
\put(24.75,3){\makebox(0,0){$+$}}
\footnotesize
\put(25.5,1){\framebox(3,3){${\bf S}_{N}^{N-L}$}}
\scriptsize \put(29,4.25){\makebox(0,0){$-1$}} \footnotesize
\Large
\put(2.5,2.5){\makebox(0,0){${\bf 0}$}}
\put(26,9.5){\makebox(0,0){${\bf 0}$}} \normalsize
\end{picture}
}
\right]
\end{equation}

\bibliographystyle{IEEEbib}
\bibliography{ref}

\end{document}